\documentclass[letterpaper,12pt,reqno]{amsart}

\usepackage{todonotes,latexsym, amssymb,amsthm, natbib,verbatim, xy}
\usepackage{todonotes,tgpagella, mathpazo, mathtools, appendix}
\xyoption{all}
\usepackage{setspace,color}
\usepackage{graphicx}

\usepackage[left=1in,top=1.125in,right=1in,bottom=1.125in]{geometry}
\usepackage{enumerate}
\usepackage[multiple]{footmisc}
\usepackage[leqno]{amsmath}
\usepackage{hyperref}
\usepackage{yhmath}
\usepackage{ifsym} 
\usepackage{ulem}
\usepackage{etoolbox}
\patchcmd{\section}{\scshape}{\bfseries}{}{}

\makeatletter
\renewcommand{\@secnumfont}{\bfseries}
\makeatother

\usepackage{tikz}
\usetikzlibrary{shapes.misc, shapes.geometric, positioning}
\newcommand*\circled[1]{\tikz[baseline=(char.base)]{
    \node[shape=circle,draw,inner sep=1pt, scale=0.4] (char) {#1};}}

\renewenvironment{quote}{%
   \list{}{%
     \leftmargin0.75cm   
     \rightmargin0.5cm
   }
   \item\relax
}
{\endlist}

\newenvironment{indquote}{
    \begin{quote}
    \fontfamily{lmtt}\selectfont\small
}
{
    \end{quote}
}

\newenvironment{proof*}[1][Proof]{\noindent{}\textbf{\textit{#1.}} }{\ \hfill $\blacksquare$ \\}

\renewenvironment{proof}[1][Proof]{\noindent{}\textbf{\textit{#1.}} }{\ \hfill $\square$ \\}

\newtheorem{theorem}{Theorem} 
\newtheorem{corollary}{Corollary}
\newtheorem{lemma}{Lemma}
\newtheorem{lemma2}{Lemma} 
\newtheorem{proposition}{Proposition} 
\newtheorem{claim}{Claim}
\theoremstyle{definition}
\newtheorem{remark}{Remark}

\newtheorem{definition}{Definition}

\def\cala{\mathcal{A}} 
  
\def\calx{\mathcal{X}}
\def\cali{\mathcal{I}}
\def\calj{\mathcal{J}}
\def\calf{\mathcal{F}}
\def\calq{\mathcal{Q}}
\def\cale{\mathcal{E}}

\def\calr{\mathcal{R}} 
\def\calv{\mathcal{V}} 
\def\calm{\mathcal{M}}

\newcommand*{\medcap}{\mathbin{\scalebox{1.25}{\ensuremath{\cap}}}}
\newcommand*{\medcup}{\mathbin{\scalebox{1.25}{\ensuremath{\cup}}}}

\newcommand{\abs}[1]{\left| #1 \right|}

\def\o{\circled{OA}}
\def\h{\circled{AT}}

\def\co{C_{\o}}
\def\aco{\widehat{C}_{\o}}

\def\ews{e}
\def\scores{\Sigma}

\def\mem{\mathbb{P}}
\def\memone{\mathbb{P}^1}
\def\memtwo{\mathbb{P}^2}
\def\memonezerothree{\mathbb{P}^{103}}

\def\bbm{\mathbb{M}}
\def\u{\uline}
\def\n{\wideparen}
\def\h{\widehat}

\begin{document}

\title[]{Informed Neutrality in Minimalist Market Design:\\ A Case Study on a Constitutional Crisis in India}

\author[]{Tayfun S\"{o}nmez \and M. Utku \"{Unver}}

\thanks{This paper replaces and subsumes \cite{sonmezunver22} and \cite{sonmezunver22b}, 
two earlier versions circulated right before and right after the judgment \textit{Janhit Abhiyan (2022)\/}, which was delivered in a 3-2 split verdict by the Supreme Court of India. 
Both S\"{o}nmez and \"{Unver} are affiliated with the Department of Economics, Boston College, 140 Commonwealth Ave, Chestnut Hill, MA, 02467. Emails: \texttt{sonmezt@bc.edu},
\texttt{unver@bc.edu}. We thank Ashwini Deshpande, Manshu Khanna and participants of the NBER Fall 2023 Market Design Working Group Meeting and the University of
Tokyo Market Design Center Special Lecture Series for their detailed comments.}

\date{First Version: October 10th, 2022. This Version: May, 2024}

\maketitle

\begin{abstract}
In a 3-2 split verdict, the Supreme Court approved the exclusion of India's socially and economically backward classes from its affirmative action measures to address economic deprivation. 
Dissenting justices, including the Chief Justice of India, protested the Majority Opinion for sanctioning ``an avowedly exclusionary and discriminatory principle.''
To justify their controversial decision, the majority justices relied on technical arguments that are categorically false. 
The confusion of the justices is due to a subtle technical aspect of the affirmative action system in India: the significance of overlaps between members of protected groups. 
Conventionally, protected classes were determined by the caste system, which meant they did not overlap. The addition of a new protected class defined by economic criteria alters this structure, 
unless it is artificially enforced. The majority justices failed to appreciate the significance of this critical change and inaccurately argued that the controversial exclusion is a technical 
necessity to provide benefits to previously unprotected members of a new class. We show that this case could have been resolved with three competing policies that each 
avoid the controversial exclusion. One of these policies aligns with the core arguments in the Majority Opinion, whereas a second aligns with those in the Dissenting Opinion.
\end{abstract}



\newpage

\hfill\textit{A dissent in a court of last resort is an appeal to the} 

\hfill\textit{brooding spirit of law, to the intelligence of a future day, when}

\hfill\textit{a later decision may possibly correct the error into which the} 

\hfill\textit{dissenting justice believes the court to have been betrayed.} {\vspace{0.5em}}

\hfill\textit{-- Charles Evans Hughes \citeyearpar{hughes1928supreme}}

\section{Introduction} \label{sec:introduction}

Affirmative action for India's socially and educationally disadvantaged classes is embedded in its Constitution within Articles 14-18, a segment known as the \textit{Equality Code\/}. 
The landmark Supreme Court judgment \textit{Indra Sawhney vs Union of India (1992)\/},\footnote{Widely known as the \textit{Mandal Commission Case\/}, 
the judgment serves as the main reference for India's reservation system and is available at \url{https://indiankanoon.org/doc/1363234/}. Last retrieved on 05/13/2024.} 
henceforth \textit{Indra Sawhney (1992)\/}, codified these provisions into a powerful positive discrimination policy called \textit{vertical reservations (VR)\/}.

Originally, the VR policy was envisioned as a reparative and compensatory mechanism to level the playing field for classes facing social marginalization based on their hereditary caste identity. 
Up to 49.5\% of positions in government jobs and institutions of higher education were set aside for Scheduled Castes (SCs), Scheduled Tribes (STs), and Other Backward Classes (OBCs). 
However, in a highly contentious reform in 2019, it was extended to a class based on a transient 
characteristic.\footnote{See, for example, the \href{https://scroll.in/article/1037199/economist-ashwini-deshpande-on-why-reservations-are-not-the-right-instrument-to-reduce-poverty}{\textit{Scroll.in\/} interview} 
``Economist Ashwini Deshpande on why reservations are not the right instrument to reduce poverty,'' last accessed 05/13/2024.}

The 103rd Constitutional Amendment,\footnote{The Amendment is officially called the \textit{Constitution (One Hundred and Third Amendment) Act, 2019\/}.} 
introduced VR protections of up to 10\% of positions for a new category known as \textit{Economically Weaker Sections (EWS)\/}. 
Eligibility for EWS is provided to financially disadvantaged individuals, but controversially, it is restricted to those who fall outside the scope of the earlier reparative and compensatory VR protections. 
Consequently, since its introduction, EWS reservations have functioned as a positive discrimination policy for a subclass of India's more privileged forward classes.

The Amendment was immediately challenged by several groups and ultimately advanced to a five-judge Constitution Bench of the Supreme Court in August 2020. 
In September 2022, the Bench announced its decision to address the following three main issues to determine whether 
the Amendment violates the basic structure of the Constitution:
\begin{enumerate}
\item Can reservations be granted solely on the basis of economic criteria?
\item Can states provide reservations in private educational institutions which do not receive government aid?
\item  Are EWS reservations constitutionally invalid for excluding  SCs, STs, and OBCs?
\end{enumerate}

In a landmark judgment \textit{Janhit Abhiyan vs Union of India (2022)\/}, henceforth \textit{Janhit Abhiyan (2022)\/}, 
the Constitution Bench announced its verdict in November 2022. The Supreme Court upheld the 103rd Constitutional Amendment, albeit in a 3-2 split verdict.\footnote{There are four separate judgments in
\textit{Janhit Abhiyan vs. Union of India (2022)\/}: a Majority Opinion by Justice Dinesh Maheshwari, representing the decision; two ``obiter dicta'' judgments by the other two majority justices, 
Justice Bela Trivedi and Justice J.B. Pardiwala, indicating their personal comments, observations, and suggestions; 
and a Dissenting Opinion by Justice Ravindra Bhat and Chief Justice of India Uday Umesh Lalit. The judgment containing all four opinions is available at the following link: 
\url{https://www.scobserver.in/wp-content/uploads/2021/10/EWS-Reservations-Judgment.pdf}.
See also the \href{https://www.scobserver.in/wp-content/uploads/2021/10/EWS-Reservations-Judgment-3-157.pdf}{Majority Opinion}, 
the \href{https://www.scobserver.in/wp-content/uploads/2021/10/EWS-Reservations-Judgment-158-181.pdf}{Obiter Dictum of Justice Bela Trivedi}, the \href{https://www.scobserver.in/wp-content/uploads/2021/10/EWS-Reservations-Judgment-182-298.pdf}{Obiter Dictum of Justice J.B. Pardiwala}, and the \href{https://www.scobserver.in/wp-content/uploads/2021/10/EWS-Reservations-Judgment-299-399.pdf}{Dissenting Opinion} 
of Justice Bhat and Chief Justice of India Lalit. All links were last accessed on 05/13/2024.}

While all five justices agreed that reservations can be granted solely on the basis of economic criteria, the two dissents— including the Chief Justice of India—strongly disagreed 
with the majority justices on the constitutionality of the exclusion of socially and educationally disadvantaged classes from the scope of EWS. 
The extent of the disagreement can be vividly seen in the opening paragraph of the Dissenting Opinion by Justice Bhat.\smallskip

\begin{indquote}
``1. I regret my inability to concur with the views expressed by the majority
opinion on the validity of the 103rd Amendment on Question No. 3, since I
feel - for reasons set out elaborately in the following opinion - that this
court has for the first time, in the seven decades of the republic, sanctioned
an avowedly exclusionary and discriminatory principle. Our Constitution
does not speak the language of exclusion. In my considered opinion, the
Amendment, by the language of exclusion, undermines the fabric of social
justice, and thereby, the basic structure.''\smallskip
\end{indquote}

The main themes of this paper include (i) the contested exclusion of the socially and educationally disadvantaged classes from the scope of the EWS,  
(ii) the technical errors in the Majority Opinion's justification for upholding this exclusion, and 
(iii) the alternative decisions that would be consistent with existing legislation on reservation policy in the absence of these errors. 
Throughout the paper, we refer to the policy that removes the controversial exclusion in the EWS as the \textit{scope-increased EWS policy\/}.

While the verdict in  \textit{Janhit Abhiyan (2022)\/}  was declared a major victory for the central government led by Prime Minister Narendra Modi,\footnote{See, for example, the
 \href{https://www.thehindu.com/news/national/ews-quota-verdict-historic-a-victory-for-pms-mission-bjp/article66108757.ece}{\textit{The Hindu\/} story} 
 ``EWS quota verdict historic, a victory for PM’s mission: BJP,'' last accessed 05/13/2024.}  
according to many,  it also sparked uproar in the country.\footnote{See, for example, the
 \href{https://www.aljazeera.com/news/2022/11/9/why-10-quota-for-economically-weak-in-india-has-caused-uproar}{\textit{Al Jazeera\/} story} 
 ``Why 10\% quota for `economically weak' in India has caused uproar,'' and the 
 \href{https://www.livelaw.in/interviews/ews-quota-is-reservation-for-the-over-represented-excludes-the-real-unrepresented-mohan-gopal-on-supreme-court-judgment-interview-213981}{\textit{LiveLaw\/} 
 interview}``EWS Quota Is Reservation For The Over-Represented, Excludes The Real Unrepresented : Mohan Gopal On Supreme Court Judgment,'' last accessed 05/13/2024.}  
The following paragraphs from an opinion piece by Anup Surendranath, a Professor of Law at the National Law University in Delhi, summarize the key 
elements of the turmoil caused by the judgment:\footnote{See \href{https://indianexpress.com/article/opinion/columns/ews-reservation-supreme-court-equality-non-discrimination-affirmative-action-8274435/}{\textit{The Indian Express\/} opinion} 
``EWS reservation: Supreme Court has not clarified tricky questions at the intersection of equality, non-discrimination, and affirmative action,'' last accessed 05/13/2024.}
\begin{quote}
``It is constitutionally perverse that the compelling need for measures to address social backwardness has 
become a justification for the exclusion of backward classes from measures to address economic deprivation.

The historical confusions in the Supreme Court’s reservation jurisprudence have come home to roost in the EWS judgment. 
Despite seven decades of constitutional adjudication on reservation, fundamental questions remain unexamined resulting in a jurisprudence that permits constitutionally perverse outcomes. 
More precisely, the constitutionally perverse outcomes resulting from the majority judgment are: India's most marginalised sections that comprise a significant proportion of India's poor stand excluded from reservation meant for the poor, 
and second, it is now far easier to provide reservation for this narrowly constructed EWS than it is to do the same for India’s most marginalised sections. 
These outcomes are fuelled by a flawed constitutional logic that does tremendous disservice to the founding constitutional agreement, social history and lived reality of India’s most vulnerable sections.''
\end{quote}

\subsection{Politically Motivated Decision or a Technical Oversight?} \label{subsec:political}

The main outrage against the judgment stems from the widespread perspective that, by upholding the controversial exclusion, it betrays the philosophy of affirmative action. 
Specifically, due to this exclusion, a member of a privileged group can secure a position over a member of a disadvantaged group, despite the privileged individual 
having a lower merit score and both individuals suffering financial disability. Naturally, this anomaly cannot be justified by either meritocracy or affirmative action. 
Consequently, many in the country, including the dissenting justices, maintain that the exclusionary clause in the Amendment violates the country's \textit{Equality Code\/}.

One prominent view is that the decision in \textit{Janhit Abhiyan (2022)\/} is politically motivated. 
Supporting this perspective, the three majority justices include Justice Pardiwala, a well-known opponent of the reservation system. 
Justice Pardiwala has previously commented that, along with corruption, reservation is one of the two factors hindering 
India's progress in the right direction.\footnote{See \href{https://caravanmagazine.in/law/ews-reservation-judgement-supreme-court-caste-bias}{\textit{Caravan\/} article} 
``EWS judgement undermines Constitutional code of equality: Legal experts, Bahujan leaders,'' last accessed 05/13/2024.}  
Reinforcing his stance against reservation, in paragraph 187 of his Obiter Dictum, Justice Pardiwala remarked:\smallskip
\begin{indquote}
``The new concept of economic criteria introduced by the impugned amendment for affirmative action may go a long way in eradicating caste-based reservation. 
It may be perceived as a first step in the process of doing away with caste-based reservation.'' \smallskip
\end{indquote}
Moreover, this perspective is echoed in paragraph 28 of Justice Trivedi's Obiter Dictum, another majority justice in \textit{Janhit Abhiyan (2022)\/}:\smallskip
\begin{indquote}
``However, \underline{at the end of seventy-five years of our independence, we need to} \underline{revisit the system of reservation in the larger interest of the society as a}
\underline{whole, as a step forward towards transformative constitutionalism}.''\smallskip
 \end{indquote} 

Thus, with two of the three majority justices expressing opposition to the reservation system, we cannot dismiss the possibility that the decision was politically motivated. 
However, in this paper, we offer an alternative perspective: a significant technical oversight by the majority justices, which we discuss in detail in Section \ref{sec:critical}, may have also contributed to their controversial decision.

To contribute to this important debate, we present an analysis of the judgment from a completely different angle. 
Our paper highlights the flaws of the Majority Opinion from a market design perspective. Following our formal analysis in Sections \ref{sec:model} and \ref{sec:3choicerules}, 
in Section \ref{sec:critical} we refute the main arguments in the Majority Opinion used to justify the exclusion of socially and educationally backward classes from the scope of EWS reservation. 
Therefore, we conclude that the majority justices fundamentally altered the structure of affirmative action in India based on false premises. In contrast, our formal analysis 
underscores the virtue of the Dissenting Opinion and demonstrates how it could have been operationalized through three alternative policies, each eliminating the controversial exclusion with its distinct normative justification.

The confusion among the majority justices largely stems from a subtle technical aspect of the country's reservation system, which we discuss in the following section.

\subsection{Implementation of VR Policy} \label{sec:implementationVR}

While the EWS reservation is also applicable to more complex scenarios, both the debates in \textit{Janhit Abhiyan (2022)\/} and the policy discussions in India exclusively focus on the most basic scenario. 
Hence, throughout the paper, we concentrate on this basic version to directly relate our analysis to Supreme Court judgments and policy discussions in India. 
In this scenario, a set of identical positions must be allocated among applicants based on both their merit scores and their eligibility for VR-protected categories.\footnote{More involved applications of 
the problem introduce two types of complexities. In addition to the primary vertical reservations, there are also secondary reservations in India known as \textit{horizontal reservations\/} \citep{sonmez/yenmez:22}. 
Furthermore, in some applications, allocation is carried out across multiple institutions, thus introducing heterogeneity in positions \citep{sonmez/yenmez:22b}. 
While our focus remains on the basic version with identical positions and the VR policy only, enabling direct relevance to the judgment \textit{Janhit Abhiyan (2022)\/} 
as well as the policy discussions on EWS reservations in India, all our main messages extend to the more complex versions of the problem as well.} 
For each VR-protected category, such as SC, ST, OBC, and EWS, a specific number of positions are exclusively reserved for the members of that category. 
In contrast, all applicants are eligible for the remaining positions, known as \textit{open-category\/} positions.

As documented in \cite{sonmez/yenmez:22}, the implementation of the VR policy is subject to the following mandates from \textit{Indra Sawhney (1992)\/}:\footnote{The terminology for the axioms formulating 
the \textit{Indra Sawhney (1992)\/} principles is from \cite{sonmez/yenmez:22}.}
\begin{enumerate}
\item \textit{Non-wastefulness\/}: Within any category, no position can remain unfilled while there is an eligible applicant who remains unassigned.
\item \textit{No Justified Envy\/}: Within any category, no individual can be awarded a position while a higher merit-score individual eligible for the position remains unassigned.\footnote{In legal language, 
this mandate is known as \textit{the principle of merit\/} for open-category positions and \textit{the principle of inter se merit\/} for VR-protected positions.}
\item \textit{Compliance with VR Protections\/}: No VR-protected position should be allocated to an individual who qualifies for an open-category position based on merit without invoking the provisions of the VR policy. 
VR-protected positions should instead be awarded to the highest merit individuals who do not qualify for an open position.
\end{enumerate}

When no individual is a member of multiple VR-protected categories, as it has always been in India, 
the mandates of \textit{Indra Sawhney (1992)\/} are uniquely satisfied by the \textit{Over-and-Above\/} (O\&A) choice rule: First, open positions are awarded to individuals with the highest merit scores, and then, 
for each VR-protected category, the reserved positions are awarded to the remaining members of the category with the highest merit scores. 
Critically, because there is no overlap between memberships of any two VR-protected categories, the method of allocating VR-protected positions across 
multiple categories under the O\&A choice rule does not affect the outcome. Because individuals do not compete for positions in multiple VR-protected categories, competitions within these categories are entirely independent.

In contrast, when individuals are eligible for positions in multiple VR-protected categories, their allocations can interfere with each other, potentially affecting the distribution of VR-protected positions. 
Specifically, without a clear specification of how EWS positions are allocated in relation to other VR-protected categories, simply expanding the scope of the 
EWS reservation to include all individuals with financial disabilities no longer results in a well-defined system under current legislation. 
This is why the removal of the controversial exclusion, i.e., the adoption of the \textit{scope-increased EWS policy\/}, fundamentally alters a key aspect of the reservation system.

At this point, it is worth highlighting a technical point concerning a major oversight in the Majority Opinion. In examining whether the exclusion renders the EWS reservation constitutionally invalid in 
\textit{Janhit Abhiyan (2022)\/}, the justices indirectly assess the constitutionality of the non-overlapping structure of VR protections. Naturally, their decision must not rely on critical assumptions 
on vertical reservations that are no longer valid under overlapping VR protections. However, as we will establish in Section \ref{sec:critical}, the majority justices exclusively relied on arguments 
hinging on the existing structure of non-overlapping VR protections, which would not apply under overlapping VR protections. In our view, this technical 
oversight is one of the main reasons for the flawed arguments in the Majority Opinion.

\subsection{Compliance with the Equality Code}

The burden of proof to repeal a constitutional amendment is highly demanding in India, requiring evidence that the amendment violates the basic structure of the constitution. 
According to the plaintiffs and the dissenting justices, the contested 103rd Amendment does so by violating the Equality Code through the exclusion.
 Most critically, as highlighted in Section \ref{subsec:political}, a financially disadvantaged member of a more privileged group can secure a position over a financially disadvantaged member of a less privileged group, 
 despite having a lower merit score. Therefore, motivated by the Dissenting Opinion, a minimum requirement to operationalize the VR policy with the new EWS category can be formulated as follows:

\begin{itemize}
\item[(4)] \textit{Compliance with the Equality Code\/}: Consider two individuals, $i$ and $j$, who both satisfy the financial criteria for EWS eligibility, where individual $i$ belongs to a forward class while individual $j$ belongs to a socially and educationally disadvantaged class and has a higher merit score than their higher-privilege peer. Then, whenever individual $i$ is awarded a position, so shall individual $j$.
\end{itemize}

\subsection{EWS Reservation Crisis as a Case Study in Minimalist Market Design} \label{sec:minimalist}

Instances highlighting 
the failure of the current system to satisfy the axiom of \textit{compliance with the Equality Code\/} were making headlines in India as early as 2019. 
For example, the headline of a July 2019 \textit{The New Indian Express\/} article reads, ``Lower cut-off than ST, uproar over EWS quota in SBI exam for clerical-level posts.''\footnote{The story is available at \url{https://www.newindianexpress.com/cities/chennai/2019/Jul/25/lower-cut-off-than-st-uproar-over-ews-quota-2009006.html}, last accessed on 05/19/2024.} 
By the summer of 2021, similar headlines had become so prevalent that we decided to work on this crisis and provide a resolution through an institution design paradigm called \textit{minimalist market design} \citep{sonmez:23}.

Under minimalist market design, there are three main tasks:
\begin{enumerate}
\item Identify the mission of the institution: What are the primary objectives of policymakers, system operators, and other stakeholders?
\item Determine whether the existing institution satisfies these objectives. If it doesn't, then there is potential for policy impact. To materialize this potential into a successful redesign, identify the ``root causes'' of the failures.
\item Address these failures by interfering only with the flawed aspects of the deficient institution.
\end{enumerate}

This paradigm was developed to embed a persuasion strategy within an institutional design framework, particularly for situations where design economists, 
as outsiders, seek to influence policymakers who often have vested interests in maintaining the status quo.

The EWS crisis was an ideal application for minimalist market design. The primary objectives of the policymakers had already been formulated in \textit{Indra Sawhney (1992)\/}, 
but for the current crisis, these objectives needed to be supplemented with the principle of \textit{compliance with the Equality Code}.
Moreover, it was also evident that the existing system did not satisfy this principle, with the root cause being the contested exclusion of socially and educationally disadvantaged classes from the scope of EWS. 
Thus, our task was to formulate an alternative to the current system that satisfied the primary objectives by addressing the root cause of the failures—specifically, by removing the exclusion.

\subsection{Three Competing Resolutions} \label{sec:threechoicerules}

This approach provided multiple potential resolutions, but one of them—the \textit{EWS-last O\&A choice rule\/}—was the ``closest'' to the current deficient system from a consequentialist perspective. 
Under this procedure, open positions are first allocated to individuals with the highest merit scores, followed by caste-based VR-protected positions allocated 
to the remaining highest-merit score individuals eligible for these positions, and finally, EWS positions are allocated to the remaining highest-merit score individuals who satisfy the financial criteria for EWS.

Importantly, assuming the exclusion does not strictly decrease the number of EWS positions that can be feasibly assigned to eligible individuals, the EWS-last O\&A choice rule yields 
a different outcome from the current deficient system if and only if the current system violates the Equality Code (Theorem \ref{thm-EWS2}). 
Moreover, whenever the existing system violates the Equality Code, the outcome of the EWS-last O\&A choice rule differs from the existing system's outcome 
by merely replacing the beneficiaries of this violation—members of privileged classes who received positions with relatively low scores—with members of disadvantaged classes 
who failed to receive positions despite having relatively higher scores. Indeed, once the controversial exclusion is removed, any choice rule that satisfies the core mandates of \textit{Indra Sawhney (1992)\/} 
results in displacing a set-wise larger set of individuals from forward classes compared to the set of affected individuals from forward classes when the current system 
is replaced with the EWS-last O\&A choice rule (Theorem \ref{thm-EWS}). Essentially, the EWS-last O\&A choice rule functions as a procedure that ``corrects'' any violations of the Equality Code 
whenever the current system involves one, but beyond these corrections, it maintains the selection of the current system.

As we discuss in depth in Section \ref{sec:critical}, the majority justices relied on two related arguments to uphold the constitutionality of the contested exclusion. 
They argued that, without excluding members of the traditional caste-based VR-protected classes, it is not possible to provide any meaningful positive discrimination to any group not covered by these earlier provisions. 
Moreover, they also argued that, without the exclusion, members of the traditional caste-based VR-protected classes would receive excessive benefits, referred to as the ``double benefits'' argument in the country. 
Theorems \ref{thm-EWS} and \ref{thm-EWS2} clearly establish that their justification is completely flawed. Indeed, the EWS-last O\&A choice rule only differs from the current system 
by eliminating scenarios that make headlines in the country, where disadvantaged classes are directly discriminated against by the system in favor of privileged classes.

Regarding the flawed ``double benefits'' argument, a key aspect of the current legislation is also illustrative. 
Under the current implementation of VR policy, it is voluntary for members of socially and educationally backward classes to forego their VR protections by simply not declaring their caste memberships. 
However, even in those instances, they remain ineligible for EWS membership, even if they meet the financial criteria for this new category. 
Thus, even if there were any merit in the ``double benefits'' argument, individuals from socially and educationally backward classes are denied any means to escape such ``excessive'' 
benefits by solely claiming EWS membership. This observation supports the perspective that the EWS reservation is conceptualized more as a ``forward caste'' 
reservation than one intended for financially disadvantaged individuals who have no access to traditional VR protections.

Based on these results, we initially felt that the EWS-last O\&A choice is the obvious and unique prescription of minimalist market design for this crisis.

While the first version of our paper was still in preparation in September 2022, the Constitution Bench of the Supreme Court explicitly identified the potential 
unconstitutionality of the contested exclusion as one of the main issues to be addressed. Over the next few weeks, we closely followed the hearings for  
\textit{Janhit Abhiyan (2022)},\footnote{We thank Manshu Khanna for alerting us about the hearings of \textit{Janhit Abhiyan (2022)\/}.} 
and based on the questions from Justices Lalit and Bhat, we realized that portraying the EWS-last O\&A choice rule as the definitive resolution for this crisis might be biased.
The questioning by Justices Lalit and Bhat highlighted that, at least according to some, the issues extended beyond the failure of the Equality Code.

In particular, one of the additional differences between the majority and dissenting justices became more concrete with the ruling. Under the \textit{Indra Sawhney (1992)\/} 
mandate of compliance with VR protections, an individual who qualifies for both an open position and a VR-protected position on merit must be assigned the open position, 
so that the VR-protected position can be awarded to an  eligible individual who is truly in need of affirmative action. This feature makes VR protections the strongest affirmative action provision in India.

As we present in Section \ref{sec:EWS-first}, reflecting their perspective that traditional caste-based VR protections represent a higher-order provision 
compared to EWS—a category based on a transient feature—the dissenting justices favored assigning an individual a position from EWS whenever they 
merit a position from both EWS and a caste-based VR-protected category (but not from the open category). We refer to this principle as \textit{respecting mobility from reparatory categories to EWS\/}.

Consider the following EWS-first O\&A choice rule: Open positions are first allocated to individuals with the highest merit scores, 
followed by EWS positions allocated to the remaining highest-merit score individuals who satisfy the financial criteria for EWS, 
and finally, caste-based VR-protected positions are allocated to the remaining highest-merit score individuals eligible for these positions. 
Once the controversial exclusion is removed, 
this choice rule is the unique one that complies with the \textit{Indra Sawhney (1992)\/} 
mandates and respects mobility from reparatory categories to EWS (Theorem \ref{thm-VR-caste}). 
Thus, the EWS-first O\&A choice rule uniquely represents the key arguments in the Dissenting Opinion.

Interestingly, abstracting away from the discussions in \textit{Janhit Abhiyan (2022)\/}, there is a third choice rule under overlapping VR protections that is more compelling than any other rule satisfying 
the \textit{Indra Sawhney (1992)\/} mandates from a technocratic perspective. Under both EWS-first O\&A and EWS-last O\&A choice rules, each category is processed  one-at-a-time sequentially.

In contrast, consider a choice rule that processes all VR-protected categories simultaneously after the open category. This allows the category assignment of an individual eligible 
for multiple VR-protected categories to be contingent on individuals with lower scores who are eligible for some of these categories. For instance, an individual eligible for both 
SC and EWS can be assigned the SC position if the next individual on the merit list is eligible only for EWS, and the EWS position if the next individual on the merit list is eligible only for SC.

In Theorem \ref{thm-mOA}, we show that the resulting procedure, which we call the meritorious O\&A choice rule, \textit{Gale dominates} \citep{gale1968} any other choice rule 
$C$ that satisfies the \textit{Indra Sawhney (1992)\/} mandates. That is, 
\begin{enumerate}
\item the meritorious  O\&A choice rule selects at least as many individuals as choice rule $C$, and 
\item  for any integer $k$, the merit-score of the $k^{\mbox{\footnotesize{th}}}$ highest merit-score individual selected under the meritorious O\&A choice rule is at least as high as 
the merit score of the  $k^{\mbox{\footnotesize{th}}}$ highest merit-score individual selected under choice rule $C$.
\end{enumerate}

Therefore, the set of individuals selected under the meritorious O\&A choice rule represents the highest merit group overall under the \textit{Indra Sawhney (1992)\/} mandates.

Our analytical results highlight the importance of maintaining \textit{informed neutrality} \citep{Li2017} between reasonable but competing ethical principles in minimalist market design, 
which is a second main theme of our paper and will be discussed in Section \ref{sec:conclusion}.

\subsection{Organization of the Rest of the Paper} 
In Section \ref{sec:model}, we introduce the model and provide preliminary analysis. 
Section \ref{sec:3choicerules} contains the main analysis, while related literature in market design is included at the end of this section. 
Moving on to Section \ref{sec:critical}, we connect our formal analysis to the 
Supreme Court judgment \textit{Janhit Abhiyan (2022)}, offering a critical examination of both Majority and Dissenting Opinions. 
We conclude our paper in Section \ref{sec:conclusion} by relating our analysis to the notion of \textit{informed neutrality\/} in minimalist market design. 
Finally, the proofs of formal results are relegated to  the Appendix.

\section{Model and Preliminary Analysis}\label{sec:model}

There is a finite set $\cali$ of individuals who are  competing for $q^{\Sigma} \in \mathbb{N}$ identical positions.
Each individual  $i \in \cali$ is in need of a single position,  and has a distinct merit score $\sigma_i \in \mathbb{R}_{+}$.
Let $\sigma = (\sigma_i)_{i\in\cali}$ denote the vector of merit scores. 
In the absence of an affirmative action policy,  individuals with higher merit scores have higher claims for a position. 
Throughout the paper, we fix the set of all individuals $\cali$, the number of positions  $q^{\Sigma}$, and the vector of merit scores $\sigma$.  

There are two types of affirmative action provisions in India: the primary vertical reservations (VR policy) and the secondary horizontal reservations (HR policy). 
The discussions in the country on the 103rd Amendment, including in the Supreme Court judgment \textit{Janhit Abhiyan (2022)\/}, 
abstract away from the secondary HR policy. Thus, to relate our analysis to policy debates in India, we also exclusively focus on the primary VR policy throughout.

\subsection{Vertical Reservations}\label{sec:ver}
Let  $\calr$ denote the set of \textbf{VR-protected categories\/}.   Given an individual $i\in\cali$, let $\rho_i \in 2^{\calr}$ denote
the (possibly empty) set of VR-protected categories she belongs as a member. Define $\mem = \big(2^{\calr}\big)^{|\cali|}$, and 
let $\rho = \big(\rho_i\big)_{i\in\cali} \in \mem$ denote the profile of \textbf{category memberships}.  
For each VR-protected category $c \in \calr$, let $\cali^c(\rho) = \{i\in\cali \, : \, c\in\rho_i\}$ denote the set of members of category $c$. 
We sometimes refer to  these individuals as the \textbf{beneficiaries\/} of VR protections at  category $c$.
Individuals who do not belong to any VR-protected category are members of  
a \textbf{general category} $g\not\in\calr$. Let  $\cali^g(\rho) =  \{i\in\cali \, : \, \rho_i=\emptyset\} 
= \cali\setminus\medcup_{c\in\calr}\cali^c(\rho)$ denote the set of individuals in the general category.

Based on the conventional structure of VR-protected categories in India, papers in the literature 
assume that no individual belongs to multiple VR-protected categories. 
Motivated by the Dissenting Opinion in \textit{Janhit Abhiyan (2022)\/},  we drop this assumption. 
We refer to VR-protected categories as  \textbf{overlapping\/} if some individuals can be members of multiple VR-protected categories,
and  as \textbf{non-overlapping\/} if each individual is a member of at most one VR-protected category.
For the latter case, let $\memone \subsetneq \mem$ denote the class of non-overlapping category membership profiles. 
That is, for each $\rho \in \memone$, 
\[ |\rho_i| \leq 1 \quad \mbox{ for any } \; i\in\cali. 
\]

For any VR-protected category $c\in \calr$, let $q^c \in \mathbb{N}$ be the number of positions that are exclusively set aside for the 
members of category $c$.\footnote{This type of protective policy is sometimes referred to as \textit{hard reserve\/} policy.} 
These provisions are referred to as \textbf{VR-protected positions at category} {\boldmath $c$}.  For any VR-protected category $c\in \calr$,
let $\cale^c(\rho)=\cali^c(\rho)$ denote the set of individuals who are \textbf{eligible for VR-protected positions at category} {\boldmath $c$}. 
The total number of VR-protected positions is no more than the number of all positions. That is, 
\[ \sum_{c\in\calr} q^c \leq q^{\Sigma}.  
\]
All individuals are eligible for the remaining  
\[ q^o=q^{\Sigma}-\sum_{c\in \calr} q^c
\]
positions, which are  referred to as \textbf{open-category}  (or \textbf{category-$\boldsymbol{o}$}) positions. 
Let $\cale^o(\rho)= \cale^o = \cali$  denote the set of individuals who are \textbf{eligible for open-category positions}.

Let  $\calv = \calr \cup \{o\}$ denote the set of \textbf{vertical categories for positions}. 
Throughout the paper, we fix  $q^v$ for any $v \in \calv$.

\subsection{Solution Concepts and Primary Axioms} \label{subsec-choicerule}
We next present the solution concepts used in our paper, and the primary axioms imposed on them.

\begin{definition}
Given a profile  $\rho \in \mem$ of category memberships and a category $v \in \calv$, a \textbf{single-category choice rule}
is a function $C^v\big(\rho;.\big): 
2^{\cali} \rightarrow 2^{\cali}$, such that, for any set of individuals $I\subseteq \cali$,
\[ C^v(\rho;I)\subseteq I \medcap \cale^v(\rho)  \quad \mbox{ and  } \quad  \abs{C^v(\rho;I)}\leq q^v.\]
\end{definition}
That is, for any set of individuals, 
a single-category choice rule selects a subset from those who are eligible, up to capacity.\footnote{Strictly speaking, the domain of a 
single-category choice rule $C^v$ is $\mem \times 2^{\cali}$, which is why its arguments includes a profile $\rho \in \mem$ of category memberships.
Hence, a single-category choice rule selects a subset of its eligible applicants, up to capacity, for each profile of category memberships and set of applicants.}   

\begin{definition}
Given a profile  $\rho \in \mem$ of category memberships, a \textbf{choice rule} is a multidimensional function 
$C(\rho;.) = \big(C^{\nu}(\rho;.)\big)_{\nu \in \calv} :   
2^{\cali} \rightarrow (2^{\cali})^{|\calv|}$ such that,  
for any  set of individuals 
$I\subseteq \cali$,
\begin{enumerate}
\item for any category $v \in \calv$,
\[ C^v(\rho;I)\subseteq I \medcap \cale^v(\rho)  \quad \mbox{ and  } \quad  \abs{C^v(\rho;I)}\leq q^v,
\]
\item for any two distinct categories $v,v'\in \calv$,
\[ C^v(\rho;I)\medcap C^{v'}(\rho;I)=\emptyset. \]
\end{enumerate}
\end{definition}
That is, a choice rule is a list of interconnected single-category choice rules for each category of positions,
where no individual is selected by more than a single category.

\begin{definition}
Fix a profile  $\rho \in \mem$ of category memberships.
For any choice rule $C(\rho;.)=\big(C^{\nu}(\rho;.)\big)_{\nu \in \calv}$,  the resulting \textbf{aggregate choice rule} 
$\widehat{C}(\rho;.): 2^{\cali} \rightarrow 2^{\cali}$ is given as,
for any  
$I\subseteq \cali$, 
\[ \widehat{C}(\rho;I)=\bigcup_{v\in \calv} C^v(\rho;I).\]
\end{definition}
For a given profile of category memberships and for any set of individuals,  
the aggregate choice rule yields the set of chosen individuals across all categories.

As it is discussed in depth in \cite{sonmez/yenmez:22}, the following three axioms are mandated in
India with the Supreme Court judgment \textit{Indra Sawhney (1992)\/}.
Throughout our  analysis, we focus on  choice rules that satisfy all three axioms. 

\begin{definition} \label{axiom:non-wastefulness} 
Given a profile  $\rho \in \mem$ of category memberships,
a choice rule $C(\rho;.)=\big(C^{\nu}(\rho;.)\big)_{\nu \in \calv}$ satisfies \textbf{non-wastefulness}  if,
for  any  
$I\subseteq \cali$, $v \in \calv$, and $j \in I$,
\[j \not\in  \widehat{C}(\rho;I) \; \mbox{ and } \;  |C^v(\rho;I)| < q^v   \quad \implies \quad    j \not\in \cale^v(\rho).
\]
\end{definition}
The first axiom requires no position to remain idle for as long as there is an eligible individual. 

\begin{definition}
Given a profile  $\rho \in \mem$ of category memberships,
a choice rule $C(\rho;.)=\big(C^{\nu}(\rho;.)\big)_{\nu \in \calv}$  satisfies \textbf{no justified envy}  if, for any 
$I\subseteq \cali$,\; $v \in \calv$,\;
$i\in C^v(\rho;I)$, and
$j \in \big(I\cap \cale^v(\rho)\big) \setminus \widehat{C}(\rho;I)$, 
\[ \sigma_i  > \sigma_j.
\]
\end{definition}
The second axiom requires that no individual receives a position at any category $v\in\calv$ at the expense of another individual who is 
both  eligible for the position and at the same time  has higher merit score.

\begin{definition} \label{def-VR}
Given a profile  $\rho \in \mem$ of category memberships, 
a choice rule $C(\rho;.)=\big(C^{\nu}(\rho;.)\big)_{\nu \in \calv}$ satisfies \textbf{compliance with VR protections}  if, for any
$I\subseteq \cali$, $c \in \calr$, and $i\in C^c(\rho;I)$,
\begin{enumerate}
\item $|C^o(\rho;I)| = q^o$, and
\item for every $j \in C^o(\rho;I)$,
\[ \sigma_j  > \sigma_i. \]
\end{enumerate}
\end{definition}
The third axiom requires that, an individual who is ``deserving'' of an open-category position 
due to her merit score, should be awarded an open-category position and not one that is VR-protected. 

\subsection{Sequential Choice Rules} \label{sec:SCH}

In this section we present \textit{sequential\/} choice rules,  a class of choice rules that plays a prominent role in most
real-life applications of reserve systems.
The name of the class captures the idea that, 
under each member of the class,  all positions in any category (including the open category) are processed in blocks 
following a given sequence of categories. 

Unless explicitly stated, we fix a profile  $\rho \in \mem$ of category memberships for the rest of this section.

Let $\Delta$ denote the set of all linear orders on  $\calv$. 
Each element of $\Delta$ represents a linear processing sequence of vertical categories, 
and referred to as an \textbf{order of precedence\/}. 

For any VR-protected category $c\in\calr$, let 
\begin{itemize}
\item $\Delta^o_c \subsetneq \Delta^o \subsetneq \Delta$ denote the set of processing sequences
where the open category is processed first and category $c$  is processed last, and
\item $\Delta^{o,c} \subsetneq \Delta^o \subsetneq \Delta$  denote the set of processing sequences
where the open category is processed first and category $c$  is processed second. 
\end{itemize}
 
Given a category $v\in\calv$, define the following single-category choice rule: 
For any set of individuals $I \subseteq \cali$,  
\textbf{category-}$\boldsymbol{v}$ \textbf{serial dictatorship}
$C^v_{sd}(\rho;.)$  selects the set of highest merit-score 
individuals in $I$ who are eligible for category $v$, up to capacity.

Fix  an order of precedence $\rhd\in\Delta$. 
Given  a set of individuals $I \subseteq \cali$, the outcome
of the sequential choice rule $C_{\rhd}(\rho;.)$  
is obtained with the following procedure.

	\begin{quote}
        \noindent{}{\bf Sequential Choice Rule} $C_{\rhd}(\rho; .) = \big(C^{\nu}_{\rhd}(\rho; .)\big)_{\nu \in \calv}$\smallskip

		\noindent{}{\bf Step 0 (Initiation)}: Let $I_0 = \emptyset$.

\noindent{}{\bf Step k} {\boldmath($k\in \{1, \ldots, |\calv|$)}: 
Let $v_k$ be the category which has the $k^{\mbox{\footnotesize{th}}}$ highest order of precedence under  $\rhd$. 
\[ C^{v_k}_{\rhd}(\rho;I) = 
C^{v_{k}}_{sd}\big(\rho; \big(I\setminus I_{k-1}\big)\medcap\cale^{v_{k}}\big)
\]
Let $I_k = I_{k-1} \medcup C^{v_k}_{\rhd}(\rho;I)$.\smallskip
        \end{quote}
Following their order of precedence  under $\rhd$, 
category-$v$ serial dictatorship $C_{sd}(\rho;.)$ is applied sequentially under 
this choice rule for each vertical category $v\in\calv$.

Let $\Delta^o\subsetneq\Delta$ denote the set of processing sequences where the open category is processed before any other category. 

The class $\Delta^o$ of order of precedences play a special role in problems
where each individual belongs to at most one VR-protected category, as has historically been the case.  
Assuming non-overlapping VR-protected categories, 
the outcome of any sequential choice rule $C_{\rhd^{o}}(\rho; .)$ is the same 
regardless of the order of precedence $\rhd^o \in \Delta^o$, an observation that fails to extend to the case of overlapping VR-protected categories. 

Given a profile $\rho \in \memone$ of non-overlapping category memberships,  define the  \textbf{Over-and-Above (O\&A)} choice rule
$C_{OA}(\rho;.) = \big(C_{OA}^{\nu}(\rho;.)\big)_{\nu\in\calv}$ as the choice rule that selects the outcome of 
a sequential choice rule $C_{\rhd^o}(\rho; .)$  for any order of precedence $\rhd^o \in \Delta^o$.

The following characterization directly follows from a more general characterization result in \cite{sonmez/yenmez:22},  and it is the starting point of our formal analysis. 

\begin{proposition} \label{prop0} Fix a  profile $\rho \in \memone$ of non-overlapping category memberships. 
Then, a choice rule $C(\rho;.)$ satisfies non-wastefulness,  no justified envy, and compliance with VR protections  if and only if $C(\rho;.) = C_{OA}(\rho;.)$.
\end{proposition}

Therefore, for as long as VR-protected categories are non-overlapping, 
the O\&A choice rule is the only choice rule that satisfies the  mandates of  \textit{Indra Sawhney (1992)}.
Due to the Dissenting Opinion in \textit{Janhit Abhiyan (2022)}, however, 
the more general version of the problem with overlapping VR protections is also important to explore. 
Therefore, we next turn our attention to the other choice rules. 

The following lemmata show that each sequential choice rule satisfies the first two axioms mandated by \textit{Indra Sawhney (1992)},
and for as long as it processes the open category before other categories, it also satisfies the third axiom. 

\begin{lemma} \label{basic-lem1} For any $\rho \in \mem$ and $\rhd\in\Delta$,  
the sequential choice rule  $C_{\rhd}(\rho;.)$ satisfies  non-wastefulness and no justified envy. 
\end{lemma}

\begin{lemma} \label{basic-lem2}
For any $\rho \in \mem$ and $\rhd\in\Delta^o$,  
the  sequential choice rule  $C_{\rhd}(\rho;.)$  satisfies  compliance with VR protections. 
\end{lemma}

The next lemma further shows that, 
any choice rule that satisfies the Supreme Court's mandates has to award the open-category positions 
to the same individuals who would receive them under  O\&A. 
Thus, any deviation from the outcome of the O\&A choice rule is due to the allocation of VR-protected positions only.

\begin{lemma}  \label{basic-lem3} Fix a profile $\rho \in \mem$  of category memberships, and a profile  $\rho' \in \memone$ of non-overlapping category memberships.
Let  $C(\rho;.)=\big(C^{\nu}(\rho;.)\big)_{\nu\in\calv}$ be any choice rule that satisfies non-wastefulness, 
no justified envy, and compliance with VR protections. Then,  for any set of individuals $I \subseteq\cali$,
\[C^o(\rho;I) = C^o_{OA}(\rho';I).
\]  
\end{lemma}

\section{Three Competing Choice Rules and Their Normative Foundations} \label{sec:3choicerules}

As we later present in detail in Section \ref{sec:critical}, there are some fundamental flaws in the Majority Opinion in \textit{Janhit Abhiyan (2022)\/} in regards
to some key technical aspects of VR policy, specifically for the case of overlapping VR protections. 
Consequently, our formal analysis in this section is based on a perspective that differs from the Majority Opinion in \textit{Janhit Abhiyan (2022)\/}. 
Instead, it is aligned with the Dissenting Opinion, which declared the exclusion of SC, ST, and OBCs from the scope of EWS as a violation of the Equality Code. 
Based on the Dissenting Opinion, this exclusion must be removed regardless of which generalization of the O\&A choice rule is adopted.

Since our presentation in this section directly pertains to the crisis regarding EWS, for much of our formal analysis, 
we assume that there is only one VR-protected category—EWS, denoted as $\ews\in\calr$—that shares common members with other VR-protected categories.\footnote{An exception to this restriction is Section \ref{sec:mOA}, 
where the analysis and the results are more general, holding regardless of the membership profile.} 

Let $\calr^0 = \calr\setminus\{\ews\}$ denote the set of traditional caste-based VR-protected categories. 
By assumption, categories in $\calr^0$ do not overlap with each other, but each of them may overlap with the category $\ews$.
Let $\memtwo$ denote the class of resulting category membership profiles. 
That is, given a category membership profile $\rho \in \memtwo$, for each $i\in\cali$, 
\[ |\rho_i|  \leq 2 \quad \mbox{ and } \quad  |\rho_i| =2 \implies e\in\rho_i.
\]
Note that, $\memone \subsetneq \memtwo \subsetneq \mem$.  

Given a profile $\rho \in \memtwo$, let
\[ \calm(\rho) = \bigcup_{c\in \calr^0} \cali^c(\rho)
\]
denote the set of beneficiaries of  traditional caste-based VR-protected categories in $\calr^0$ representing historically \textbf{marginalized} groups.

Given a profile $\rho \in \memtwo$,  using the terminology in India, 
we refer to individuals in 
\[ \calf(\rho) = \cali\setminus\calm(\rho)
\] 
as members of \textbf{forward} classes. 

Under the controversial Constitutional Amendment, two criteria govern eligibility for EWS membership. 
The first criterion concerns financial status. Let 
\[\calj \subseteq \cali
\]
represent the set of individuals meeting this criterion, fixed throughout the rest of our analysis. 

The second criterion is exclusionary. Eligibility for VR-protected positions in any caste-based category in $\calr^0$ rules out eligibility for EWS. 
Thus, if $\rho \in \memone$ represents  a profile of non-overlapping category memberships under the current legislation, then 
the set of EWS members is given as 
\[ \cali^e(\rho)  = \calj \setminus  \calm(\rho) = \calj \medcap \calf(\rho). 
\]
Let $\memonezerothree \subseteq \memone$ denote the resulting class of  category membership profiles corresponding 
to the current legislation.

What makes the 103rd Constitutional Amendment and the judgment \textit{Janhit Abhiyan (2022)\/}, which upheld it, controversial, is the second criterion
that excludes members of the set $\calj \medcap \calm(\rho)$ from the scope of EWS, even though they all satisfy the financial criterion. 
According to the dissenting justices, this exclusionary criterion violates the Equality Code. 
The essence of their argument for their position can be seen in paragraph 80 of the Dissenting Opinion:\smallskip

\begin{indquote} 
``I am of the opinion that the application of the doctrine classification
differentiating the poorest segments of the society, as one segment (i.e., the
forward classes) not being beneficiaries of reservation, and the other, the
poorest, who are subjected to additional disabilities due to caste
stigmatization or social barrier based discrimination - the latter being
justifiably kept out of the new reservation benefit, is an exercise in deluding
ourselves that those getting social and educational backwardness based
reservations are somehow more fortunate. This classification is plainly
contrary to the essence of equal opportunity.''\smallskip
\end{indquote}

In our interpretation, this objection largely stems from the failure of the following social justice principle under the current  O\&A choice rule due to the controversial exclusion.  

\begin{definition}\label{axiom:equality-code} 
Given a profile of category memberships $\rho \in \mem$,
a  choice rule $C(\rho;.)=\big(C^{\nu}(\rho;.)\big)_{\nu \in \calv}$ \textbf{complies with the Equality Code} for the set of applicants 
$I \subseteq \cali$, if  for any two applicants $i \in I \cap \calj \cap \calf(\rho)$  and  $j \in I \cap \calj \cap \calm(\rho)$, 
\[ i \in \widehat{C}(\rho;I) \; \mbox{ and }  \; \sigma_j > \sigma_i \quad \implies \quad j \in \widehat{C}(\rho;I). 
\]

A choice rule $C(\rho;.)=\big(C^{\nu}(\rho;.)\big)_{\nu \in \calv}$ \textbf{complies with the Equality Code} if it complies with the Equality Code for each
set of applicants. 
\end{definition}  

In words, between a member of the forward classes and a member of the backward classes, each of whom satisfies the financial criterion, 
if the member of the forward classes is awarded a position while the member of the backward classes has an even higher merit score, then the member of the backward classes must also be awarded a position.  
Failure of the axiom would mean that, even though both individuals satisfy the financial criterion, 
the individual with higher privilege receives a position at the expense of the individual with lower privilege, despite having a lower merit score.

Indeed, together with \textit{no justified envy\/}---one of the core mandates of \textit{Indra Sawhney (1992)\/}---removal of the controversial exclusion automatically addresses any violation of the Equality Code. 
Before stating this result,  we next formulate an operator $\wideparen{\phantom{\rho}}$ on category membership profiles in $\memonezerothree$ 
which removes the contested exclusion. 

Given a profile  $\rho \in \memonezerothree$, its \textbf{EWS scope-increased profile} $\wideparen{\rho} = \big(\wideparen{\rho}_i\big)_{i\in\cali} \in \memtwo$
is such that, 
\begin{enumerate}
\item $\wideparen{\rho}_i = \rho_i \cup \{\ews\}$ \; for any $i\in \calj$, and 
\item $\wideparen{\rho}_i  =  \rho_i$ \; for any $i\in\cali\setminus \calj$. \smallskip
\end{enumerate}

Given a profile $\rho \in \memonezerothree$ that corresponds to the current legislation, 
its EWS scope-increased profile $\wideparen{\rho}$ grants EWS membership to each individual in the set $\calj \medcap \calm(\rho)$
in addition to their existing memberships of categories within $\calr^0$.
Since each individual in $\calj \medcap \calm(\rho)$ is already a member of a category in $\calr^0$, 
we have  $\wideparen{\rho} \in \memtwo$, and
consequently, the O\&A choice rule is no longer well defined under the EWS scope-increased profile $\wideparen{\rho}$ of category memberships. 

\begin{proposition} \label{prop:NJE-EqualityCode}
Let $\rho \in \memonezerothree$  be a profile of non-overlapping category memberships that represents the current legislation,
and $\wideparen{\rho} \in \memtwo$ be its EWS scope-increased profile. Suppose choice rule $C(\wideparen{\rho};.)$ satisfies no justified envy. 
Then, choice rule $C(\wideparen{\rho};.)$ complies with the Equality Code. 
\end{proposition}

Thus, motivated  by the Dissenting Opinion and Proposition \ref{prop:NJE-EqualityCode}, 
in the remainder of our formal analysis, we focus on choice rules that satisfy the \textit{Indra Sawhney (1992)\/} mandates and under which the controversial exclusion is removed from the scope of EWS.

Throughout sections \ref{sec:EWS-last} and \ref{sec:EWS-first}, we fix a profile  $\rho \in \memonezerothree$ of non-overlapping category memberships that corresponds to the current legislation, 
and thereby also fixing its EWS scope-increased profile $\wideparen{\rho} \in \memtwo$.  

\subsection{The Case for the EWS-last VR Processing Policy} \label{sec:EWS-last}

We first consider a policy that processes the EWS category after all other categories and establish that, from a consequentialist viewpoint, 
this policy minimizes interference with the existing Constitutional Amendment while avoiding a violation of the Equality Code. 
As articulated in paragraph 38.2 of the Majority Opinion in \textit{Janhit Abhiyan (2022)\/}, the significance of minimizing interference 
is directly tied to the doctrine of ``Separation of Powers'' in cases involving Constitutional Amendments.\smallskip

\begin{indquote}
``The reason for minimal interference by this Court in the constitutional Amendments is not far to seek. In our constitutional set-up
of parliamentary democracy, even when the power of judicial review is an
essential feature and thereby an immutable part of the basic structure of
the Constitution, the power to amend the Constitution, vested in the
Parliament in terms of Article 368, is equally an inherent part of the basic
structure of the Constitution. Both these powers, of amending the
Constitution (by Parliament) and of judicial review (by Constitutional
Court) are subject to their own limitations. The interplay of amending
powers of the Parliament and judicial review by the Constitutional Court
over such exercise of amending powers may appear a little bit complex
but ultimately leads towards strengthening the constitutional value of
separation of powers. This synergy of separation is the strength of our Constitution.''\smallskip
\end{indquote} 

To operationalize this policy, we consider a sequential choice rule where 
\begin{enumerate}
\item the original category membership profile $\rho \in \memonezerothree$ is replaced with its EWS scope-increased profile $\wideparen{\rho}$, and
\item the order of precedence is such that  the open category is processed first and EWS is processed last.  
\end{enumerate}

The resulting choice rule differs from the current practice in two ways: First, parallel with the Dissenting Opinion in  \textit{Janhit Abhiyan (2022)\/}, 
it removes the controversial exclusion from the scope of the EWS category. Secondly, it processes 
the scope-extended EWS category after all other categories. We refer to the second, more subtle aspect of this policy as the \textbf{EWS-last VR processing policy}. 

Let $\uline{\rhd} \in \Delta^o_{\ews}$, i.e., the order of precedence $\uline{\rhd}$ orders the open category first and category $\ews$ last. 
We refer to the sequential choice rule $C_{\uline{\rhd}}\big(\wideparen{\rho};.\big)$ as the \textbf{EWS-last Over-and-Above} (EWS-last O\&A) choice rule.\footnote{Here, 
the outcome of the resulting choice rule is independent of which specific order of precedence is selected from $ \Delta^o_{\ews}$,  and therefore the formulation corresponds to a unique choice rule.} 

Without the exclusion, the sole beneficiaries of the EWS category are the financially disadvantaged members of the forward classes. 
Whenever the O\&A choice rule fails to comply with the Equality Code for a given set of individuals, by definition, a financially disadvantaged member of a forward class receives a position at the expense 
of a financially disadvantaged member of a marginalized group who, ironically, has a higher merit score than their higher-privilege peer. 
Naturally, any policy that addresses this anomaly will necessarily be detrimental to the financially disadvantaged members of the forward classes who receive unwarranted advantage under the current system. 
Our first result establishes that the EWS-last O\&A choice rule—which complies with the Equality Code by Lemma \ref{basic-lem1} and Proposition \ref {prop:NJE-EqualityCode}—minimizes this detrimental effect.

\begin{theorem} \label{thm-EWS} 
Let $\rho \in \memonezerothree$  be a profile of non-overlapping category memberships that represents the current legislation,
and $\wideparen{\rho} \in \memtwo$ be its EWS scope-increased profile.
Let $C\big(\wideparen{\rho};.\big)$ be any choice rule that satisfies non-wastefulness, no justified envy, and  compliance with VR protections.
Then, for any $I \subseteq \cali$, 
\[ \underbrace{\widehat{C}_{OA}\big(\rho;I\big) \setminus \widehat{C}_{\uline{\rhd}}\big(\wideparen{\rho};I\big)}_{\subseteq  \calf(\rho) \cap \calj}
\subseteq \underbrace{\widehat{C}_{OA}\big(\rho;I\big) \setminus \widehat{C}\big(\wideparen{\rho};I\big)}_{\subseteq  \calf(\rho) \cap \calj}.
\]
\end{theorem}

Theorem \ref{thm-EWS} establishes that, among all choice rules that meet the core mandates of \textit{Indra Sawhney (1992)\/} under the scope-increased EWS membership profile, 
the EWS-last O\&A choice rule $C_{\uline{\rhd}}\big(\wideparen{\rho};.\big)$ minimally interferes with the
existing O\&A choice rule $C_{OA}\big(\rho;.\big)$ in terms of its detrimental effect on the financially disadvantaged members of forward classes. 
Indeed, each individual chosen by the O\&A choice rule 
but not by the EWS-last O\&A choice rule would remain unselected 
by any choice rule that meets the core mandates of \textit{Indra Sawhney (1992)}.

A next natural question is the following. What if the O\&A choice rule complies with the Equality Code for a given set of individuals? 
In light of the ``minimal interference'' justification given above,  is it also the case  that the outcome of the  EWS-last O\&A choice rule
is the same as the outcome of the O\&A choice rule in such instances? 

As it turns out, this is not always the case because, regardless of the choice rule, the controversial exclusion may sometimes strictly decrease the number of 
EWS positions that can be feasibly filled. However, aside from this potential anomaly caused by the controversial exclusion, the outcome equivalence between the O\&A choice rule and the 
EWS-last O\&A choice rule does hold. This further reinforces our interpretation of ``minimal interference.''

In our next result, we assume there is always an over-demand for EWS positions, so the controversial exclusion does not result in 
discarding some of these positions, in addition to violating the Equality Code.

\begin{theorem} \label{thm-EWS2}
Let $\rho \in \memonezerothree$  be a profile of non-overlapping category memberships that represents the current legislation,
and $\wideparen{\rho} \in \memtwo$ be its EWS scope-increased profile.
Let the set of applicants $I \subseteq \cali$, the capacity vector $q = \big(q^v\big)_{v\in\calv}$, and the membership profile $\rho$ be such that  
\[ \big|I \cap \calj \cap  \calf(\rho)\big| \geq q^o + q^e.  
\]
Then, 
\[ \widehat{C}_{\uline{\rhd}}\big(\wideparen{\rho};I\big) = \widehat{C}_{OA}\big(\rho;I\big)
\]
if and only if the choice rule $C_{OA}(\rho;.)$ complies with the Equality Code for $I$.
\end{theorem}

\subsubsection{Conditional Membership for the New Members under the EWS-last VR Processing Policy} \label{sec:conditional}

It is worthwhile to highlight the important role the specific order of precedence $\uline{\rhd} \in \Delta^o_{\ews}$ plays 
under the sequential choice rule $C_{\uline{\rhd}}\big(\wideparen{\rho};.\big)$.   Under this order of precedence, 
\begin{enumerate}
\item positions in the open category are allocated prior to positions in any other category,
\item but more critically, the positions  in category $\ews$ are processed after the positions in all other categories. 
\end{enumerate}

This selection has a key implication on the nature of benefits the extra category-$\ews$ membership given to members 
of the set  $\calj \medcap \calm(\rho)$  under EWS scope-increased category memberships in $\wideparen{\rho}$.
More specifically, the potential benefits of this new membership become largely diminished for individuals in $\calj \medcap \calm(\rho)$ under the order of precedence  
$\uline{\rhd}$, and the benefits are realized only if there would be a violation of the Equality Code in the absence of their category-$\ews$
membership.\footnote{In contrast, the potential benefits would always fully kick in 
under an alternative order of precedence $\overline{\rhd} \in \Delta^{o,\ews}$, where category $\ews$ positions are allocated
immediately after the open-category positions. See Section \ref{sec:EWS-first} for a justification for this alternative policy.} 
That is because, since positions in all other categories are allocated prior to positions in category $\ews$ under the order of precedence $\uline{\rhd}$, 
relatively lower merit-score members of other VR-protected categories remain in competition for allocation of category-$\ews$ positions. 
As such, if there were no violation of the Equality Code in the absence of the extra memberships provided under $\wideparen{\rho}$, 
then all category-$\ews$ positions would be awarded to original members of this category under $\rho$. 
Therefore, the additional membership provided to members of the set $\calj \medcap \calm(\rho)$ under the sequential choice rule $C_{\uline{\rhd}}\big(\wideparen{\rho};.\big)$ 
can be interpreted as ``conditional membership,'' activating only when necessary to avoid a violation of the Equality Code.

Crucially, in Section \ref{sec:doublebenefits}, we establish that this aspect of the EWS-last VR processing policy directly contradicts two primary 
arguments outlined in the Majority Opinion of \textit{Janhit Abhiyan (2022)\/}, which served as the sole rationale for upholding the contentious exclusion in the Amendment:
\begin{enumerate}
\item \textbf{Necessity.} The exclusion of beneficiaries from caste-based VR-protected categories from EWS is deemed essential to ensure benefits are directed towards groups ineligible for these provisions.
\item \textbf{Double Benefits.} Without such exclusion, beneficiaries from caste-based VR-protected categories would receive excessive advantage.
\end{enumerate}

We conclude this section by highlighting another issue with the  the majority justices' (flawed) ``double benefits'' argument.  

In India, the members of caste-based categories in $\calr^0$ are eligible to compete for VR-protected positions set aside for them, 
but they are not required to do so as declaring these memberships is voluntary. However, a member of a caste-based category who satisfies the financial criterion—i.e., a member of the 
set $\calj \medcap \calm(\rho)$—cannot secure a membership for EWS even if they do not declare their membership for category $c$. Thus, not only does the controversial Amendment exclude beneficiaries of 
categories in $\calr^0$ from the scope of EWS, but it also continues to exclude them even when they give up their VR protections for categories in $\calr^0$.

Thus, even if the ``double benefits'' argument of the majority justices were to have merit (which it doesn't), members of the set $\calj \medcap \calm(\rho)$ are not provided with the means to claim EWS membership by 
relinquishing their VR protections from their caste-based categories.

\subsection{The Case for the EWS-first VR Processing Policy}  \label{sec:EWS-first}

As a second policy  that also deserves serious consideration under the scope-extended EWS category, 
we next consider one that processes EWS category immediately after the open category. 
We refer to this policy as  the \textbf{EWS-first VR processing policy}.  

Let $\overline{\rhd} \in \Delta^{o, \ews}$, i.e., the order of precedence $\overline{\rhd}$ orders the open category first and category $\ews$ second. 
We refer to the sequential choice rule $C_{\overline{\rhd}}\big(\wideparen{\rho};.\big)$ as the \textbf{EWS-first Over-and-Above} (EWS-first O\&A) choice rule.\footnote{As in the 
case of the EWS-last Over-and-Above choice rule, the outcome of the resulting choice rule is independent of which order of precedence is picked from $ \Delta^{o,\ews}$, 
and therefore the formulation corresponds to a unique choice rule.}

\subsubsection{Maintaining the Elevated Status of Caste-Based VR Protections}

As we have emphasized before, the VR policy is intended as the strongest form of positive discrimination policy in India. 
The reason for such a powerful policy is historical. Originally, the VR policy was intended for members of Scheduled Castes, the official term for Dalits or ``untouchables,'' 
who have endured millennia-long systematic injustice due to their lowest status under the caste system, and Scheduled Tribes, the official term for the indigenous ethnic groups of India, 
who were both physically and socially isolated from the rest of society. With the \textit{Indra Sawhney (1992)\/} ruling, the VR policy was extended to members of Other Backward Classes 
who historically engaged in various marginal occupations assigned to them by society to serve castes higher in the caste hierarchy. 
The strength of the VR protection policy reflects the strong desire in Indian society to acknowledge and reverse the excessively disadvantaged status of these communities.

Besides the significant fraction of positions reserved under the VR policy, what other aspects of this policy make it especially powerful? 
The answer to this question lies in the following two technical aspects of this policy.

First of all, positions set aside for a VR-protected category are exclusive to its beneficiaries, even if it means that some of these positions remain 
unassigned.\footnote{These unassigned units are typically carried over to the next allocation period to be added to the VR-protected positions for the same category.}

But more importantly, the landmark Supreme Court judgment \textit{Indra Sawhney (1992)\/} explicitly mandated that these positions are not to be used for members of 
the VR-protected category who could already receive a unit from the open category without invoking the benefits of the VR policy.
Thus, the VR-protected positions are explicitly directed to those who could not receive an open position. It is this second aspect of the VR policy that makes it especially powerful.

In our formal analysis, this powerful mandate is captured by the axiom \textit{compliance with VR protections\/}. Naturally, the justices did not introduce formal axioms in their judgment. 
To describe this mandate in prose, the justices use the following (somewhat misguided) terminology:

Legal language does not differentiate between categories of individuals and categories of positions in India. 
For VR-protected categories such as SC, this practice is somewhat benign because positions for category SC are exclusive to members of SC. 
However, the ``general category'' for individuals and the ``open category'' for positions are also used synonymously in the country. 
That is, legal documents regularly speak of ``open-category individuals'' or ``general-category positions''. 
Unlike for VR-protected categories, this practice is not so benign since open category positions are not exclusive to individuals from the general category.

The legal concept of \textbf{migration} (also called \textbf{mobility}) arises as a consequence of this misguided convention. 
When a member of a VR-protected category (e.g., SC) receives an open category position based on merit, they are considered to have migrated to the general/open category. 
Without this migration, a member of SC would receive a position from the general/open category, contradicting the synonymous use of  ``general'' and ``open'' category terms. 
Using this terminology, a key aspect that makes VR protections an especially strong affirmative action provision is the benefit of migration/mobility it grants from a VR-protected category to the open/general category.

A key position on mobility/migration taken in the Dissenting Opinion of \textit{Janhit Abhiyan (2022)} pertains to the following question: 
What happens if a financially disadvantaged member of a caste-based VR category (say SC) does not merit a position from the open category, but they merit a VR-protected position both from EWS and SC? 
Should the principle that requires the allocation of open-category positions before SC positions also be applied between EWS positions and SC positions? 
Equivalently, should an SC position be awarded to a financially disadvantaged member of SC who already merits an EWS position or not?
 
The third point of the following discussion from paragraph 191 of Justice Bhat's Dissenting Opinion answers this question in favor of awarding 
members of caste-based categories with the benefit of migration/mobility to the EWS category whenever they merit a position in the EWS category (but not the open category). \smallskip
 
 \begin{indquote}
``The exclusionary clause operates in an utterly arbitrary manner.
Firstly, it ``others'' those subjected to socially questionable, and outlawed
practices -- though they are amongst the poorest sections of society.
Secondly, for the purpose of the new reservations, the exclusion operates
against the socially disadvantaged classes and castes, absolutely, by
confining them within their allocated reservation quotas (15\% for SCs,
7.5\% for STs, etc.). Thirdly, it denies the chance of mobility from the
reserved quota (based on past discrimination) to a reservation benefit based
only on economic deprivation.''\smallskip
 \end{indquote}

Equivalently, the dissenting justices are in favor of allocating EWS positions before caste-based VR-protected positions, 
so that a caste-based VR-protected position is not allocated to an individual who merits an EWS position: 
 
These observations motivate the following normative principle. 

\begin{definition} \label{def-VR-caste}
Given  a profile of category memberships $\rho \in \mem$, 
a choice rule $C(\rho;.)=\big(C^{\nu}(\rho;.)\big)_{\nu \in \calv}$  \textbf{respects mobility from reparatory categories to EWS}  if, for any 
$I\subseteq \cali$, $c \in \calr^0$, and $i\in C^c(\rho;I) \medcap \cale^{\ews}(\rho)$,
\begin{enumerate}
\item $|C^{\ews}(\rho;I)| = q^{\ews}$, and
\item for every $j \in C^{\ews}(\rho;I)$,
\[ \sigma_j  > \sigma_i. \]
\end{enumerate}
\end{definition}

Our next result shows that  EWS-first O\&A choice rule uniquely represents the position of the dissenting justices:   

\begin{theorem} \label{thm-VR-caste}
Let $\rho \in \memonezerothree$  be a profile of non-overlapping category memberships that represents the current legislation,
and $\wideparen{\rho} \in \memtwo$ be its EWS scope-increased profile.
Fix an order of precedence $\overline{\rhd} \in \Delta^{o,\ews}$.  
Then, a choice rule $C\big(\wideparen{\rho};.\big)$  respects mobility from reparatory categories to EWS  and it  satisfies non-wastefulness, 
no justified envy, and compliance with VR protections  if and only if $C\big(\wideparen{\rho};.\big) = C_{\overline{\rhd}}\big(\wideparen{\rho};.\big)$.
\end{theorem}

\subsection{The Case for the Simultaneous VR Processing Policy} \label{sec:mOA}

So far we have established that the EWS-first O\&A choice rule, 
aligns with the normative arguments presented by the dissenting justices, whereas EWS-last O\&A choice rule aligns with the normative arguments of the majority justices once their flawed "technical claims" are rectified.
Although these choice rules represent natural alternatives to the current O\&A choice rule with the scope-restricted EWS, they are not the sole compelling options.

It is well-established in the literature that when two sequential choice rules only differ in the relative order of precedence of a single category $c \in \calr$ 
compared to all other categories, the beneficiaries of category $c$ are weakly worse off  in the version where category $c$ is processed earlier \citep{dur/kominers/pathak/sonmez18, pathak/sonmez/unver/yenmez:20}. 
This perspective suggests the need to consider choice rules that do not favor beneficiaries of one category over another, advocating for joint processing of 
VR-protected categories—a policy we call \textbf{simultaneous VR processing policy}.

This normative stance finds support in the unanimous decision by all justices of \textit{Janhit Abhiyan (2022)\/} that VR protections can be granted solely based on economic criteria. 
Considering that the justices did not differentiate between EWS reservations and historical caste-based reservations, 
it is hardly unreasonable to expect the implementation to maintain neutrality between beneficiaries of various VR-protected categories.

In this section, we drop the assumption that EWS is the only category that overlaps with other VR-protected categories.
We next introduce and analyze the \textit{meritorious Over-and-Above} (meritorious O\&A) choice rule for this more general environment
as a choice rule that implements the simultaneous VR processing policy.  

Fix a   profile of category memberships $\rho \in \mem$. The following pair of auxiliary definitions simplify 
the formulation of  the  meritorious O\&A choice rule. 

\begin{definition}
Let  $\beta :   2^{\cali} \rightarrow \mathbb{N}$ denote the \textbf{VR-maximality function} that
gives the maximum number of VR-protected positions that can be awarded to eligible individuals in $I$, for any set of individuals $I\subseteq \cali$.  
\end{definition}

When VR protections are non-overlapping, for any set of individuals $I \subseteq \cali$, this number is given as 
	\[ 
		\beta(I)=\sum_{c \in \calr} \min\big \{ \big|I \medcap \cale^c(\rho) \big|, q^c\big\}.
	\]

In the general case with overlapping VR protections, this number can be  found in polynomial time  using the renowned \textit{Hungarian maximum matching algorithm\/},
initially presented in \cite{kuhn55}, building upon the earlier contributions of Hungarian mathematicians D\'{e}nes K\"{o}nig and Jen\"{o} Egerv\'{a}ry.

\begin{definition} For any set of individuals $I\subseteq \cali$ and an 
individual $i \in \cali \setminus I$,  \textbf{individual $\boldsymbol{i}$ increases the VR-utilization of $\boldsymbol{I}$} 
(upon joining individuals in set $I$) if, 
	\[
		\beta(I\cup \{i\})=\beta(I)+1.
	\] 
\end{definition}
	
Given   a profile of category memberships $\rho \in \mem$ and  a set of individuals $I \subseteq \cali$, the outcome of the meritorious O\&A choice rule $\co (\rho;.)$ is 
obtained in two steps  with the following procedure.

\begin{quote} 
\textbf{Meritorious O\&A Choice Rule} $\co (\rho;.)=\big(\co^\nu (\rho;.) \big)_{\nu \in \calv}$\smallskip

	\noindent{}{\bf Step 1.} Open category is processed. 
	Choose the highest   merit score individuals in $I$  one at a time until $\min\{|I|,q^o\}$ individuals are chosen. Let $\co^o(\rho;I)$ be the set of chosen individuals.\smallskip
	
	\noindent{}{\bf Step 2.} All VR-protected categories are processed together. \\
	Let $J=I\setminus \co^o(\rho;I)$ be the set of remaining individuals in $I$. \smallskip

	\begin{quote}
		\noindent{}{\bf Step 2.0 (Initiation)}: Let $J_0 = \emptyset$. 
	
		\noindent{}{\bf Step 2.k} {\boldmath($k\in \{1, \ldots, \sum_{c\in \calr}  q^c\}$)}:
		Assuming such an individual exists,
		choose the highest  merit-score individual in $ J \setminus J_{k-1} $	who increases the VR-utilization of $J_{k-1}$.
		Denote this individual by $j_k$ and  let $J_k=J_{k-1}\cup \{j_k\}$.
		If no such individual exists, then end the process.
	\end{quote}
	\noindent{}For any $c \in \calr$, let $\co^c(\rho;I)$ be the set of individuals who each received a category-$c$ position in this step. \smallskip
\end{quote}

\begin{remark}
While the set of individuals $\aco (\rho;.)$ who are chosen under the above-given procedure is uniquely defined, 
there may be multiple ways to assign some individuals to their VR-protected categories. 
This potential multiplicity is benign for our  analysis. 
\end{remark}

The meritorious O\&A choice rule satisfies the mandates of \textit{Indra Sawhney (1992)\/} regardless of the category membership profile. 
Furthermore, given the current structure of caste-based category memberships in India, it complies with the Equality Code, similar to the EWS-first and EWS-last O\&A choice rules, 
provided the controversial exclusion is removed from EWS membership.

\begin{proposition} \label{prop-mOA} 
For any $\rho \in \mem$, 	
the meritorious O\&A choice rule $\co(\rho;.)$ satisfies non-wastefulness, no justified envy,  and compliance with VR protections.
Moreover, for any profile $\rho \in \memonezerothree$ of non-overlapping category memberships that represents the current legislation
and its EWS scope-increased profile $\wideparen{\rho} \in \memtwo$, the meritorious O\&A choice rule $\co(\wideparen{\rho};.)$  
also complies with the Equality Code. 
\end{proposition}

We need the following terminology to present our next result. 

\begin{definition}[\cite{gale1968}]
Let members of two sets of individuals $I=\{i_1,\ldots,i_{|I|}\}, J=\{j_1,\ldots,j_{|J|}\} \subseteq \cali$ be each enumerated such that
the higher the merit score of an individual is the lower index number she has. 
Then, the set of individuals $I$ \textbf{Gale dominates} the set of individuals $J$ if,  
\begin{enumerate}
\item $|I|\geq|J|$, \; and\\ 
\item for each $\ell\in\{1,\ldots,|J|\}$, 
\[ \sigma_{i_{\ell}} \geq \sigma_{j_{\ell}}.
\] 
\end{enumerate}
\end{definition}
When a set of individuals $I$ Gale dominates another set of individuals $J$,
\begin{enumerate}
\item there are at least as many individuals who are admitted under $I$ as under $J$, and
\item the highest merit-score individual in $I$ is at least as meritorious as the highest merit-score individual in $J$, 
the second highest merit-score individual in $I$ is at least  as meritorious as the second highest merit-score individual in $J$, and so on. 
\end{enumerate}

Our final result establishes that while the uniqueness of a choice rule that satisfies the mandates of the Supreme Court's \textit{Indra Sawhney (1992)\/} 
is lost under overlapping VR protections, of all such choice rules, there still exists one that promotes meritocracy better than any other.
Thus, Theorem \ref{thm-mOA}   justifies the naming of the meritorious O\&A choice rule. 

\begin{theorem} \label{thm-mOA} 	
Let  $C(\rho;.)$ be any choice rule that satisfies non-wastefulness, no justified envy,  and compliance with VR protections. 
Then for any set of individuals $I \subseteq \cali$, the set of individuals $\aco(\rho;I)$ admitted by the meritorious O\&A choice rule
Gale dominates the set of individuals $\widehat{C}(\rho;I)$  admitted under choice rule  $C(\rho;.)$.
\end{theorem}

\subsection{Related Literature}

Our paper is an exercise in \textit{minimalist market design} \citep{sonmez:23}, a joint paradigm in research and policy. 
Methodologically, our paper benefits from the insights of \cite{moulin_book91,moulin_fair2004}, 
\cite{thomson:01}, \cite{Li2017}, \cite{Schummer2019},  \cite{Jackson2019}, and \cite{hitzig2018, hitzig2020}. 

Four papers that are especially related to our  formal 
analysis are  \cite{kominers/sonmez:16}, \cite{sonmez/yenmez:22}, \cite{dur/kominers/pathak/sonmez18}, and \cite{pathak/rees-jones/sonmez:20}.

Our model builds on  \cite{kominers/sonmez:16}, which introduces a general model with slot-specific priorities. 
\cite{sonmez/yenmez:22} formulates the Indian reservation system with both the primary vertical and the secondary horizontal reservations, 
and shows that there is a unique mechanism which satisfies the mandates of the Supreme Court judgments \textit{Indra Sawhney (1992)}
and  \textit{Saurav Yadav vs The State Of Uttar Pradesh (2020)\/}.
The  choice rule characterized in this paper, 
namely the \textit{two-step minimum guarantee\/} choice rule, is endorsed by the Supreme Court in their judgment 
\textit{Saurav Yadav (2020)\/}.\footnote{While this mechanism itself is merely endorsed by the Supreme Court, it is de facto
mandated due to the uniqueness result by  \cite{sonmez/yenmez:22}.}
Our generalization is key to analyze the third issue questioned  in  \textit{Janhit Abhiyan (2022)\/}, namely the constitutionality of the exclusion 
of socially and economically backward classes from the scope of EWS. 
Critically, in the absence of additional normative criteria, the uniqueness result by \cite{sonmez/yenmez:22}  no longer holds under our generalization. 
And as we have thoroughly discussed in Sections  \ref{sec:EWS-last}-\ref{sec:mOA}, this observation has major policy 
implications in relation to the current debates in the country in relation to the controversial Amendment.  

Practical and policy relevance of the processing sequence of categories under sequential choice rules
was first documented  in \cite{dur/kominers/pathak/sonmez18} 
for allocation of seats at Boston Public Schools (BPS) between years 1999-2013. 
As a compromise between a faction which demanded neighborhood assignment and another which demanded 
more comprehensive school choice, in 1999 leadership at BPS announced that neighborhood students will 
receive preferential treatment in half of the seats at each public school. 
This policy was referred to as \textit{walk-zone priority\/}. However, processing the walk-zone seats prior to remaining ones 
effectively negated this policy until 2013.  After discovering  that their policy was superfluous and misleading, 
the walk zone policy was abandoned at BPS altogether.

A related phenomenon for allocation of H1-B visas in the US  is presented  in \cite{pathak/rees-jones/sonmez:20}. 
With the  \textit{H-1B Visa Reform Act of 2004\/}, the US Congress reduced the number of annual H1-B visas from 195,000 to
65,000, but granted an exemption of 20,000 units for holders of advanced degrees. 
Reflecting a number of purely logistical constraints, the procedure that is used to implement this act 
was changed a few times over the years,  
although each time with significant (but likely unrealized) distributional implications.  
In response to the former President Trump’s \textit{Buy American and Hire American Executive Order in 2017\/}, however, 
the procedure was reformed by the U.S. Department of Homeland Security with an explicit objective of 
increasing the number of awards to recipients with advanced degrees. 
The latest reform, which led to an adoption of a new visa allocation rule for US Fiscal Year 2020, simply involved 
a reversal of processing sequence of advanced degree visas and general category visas. 

Our main formal results are Theorems \ref{thm-EWS} through \ref{thm-mOA}. The conceptual formulations of Theorems \ref{thm-EWS} and \ref{thm-EWS2}, 
as well as their proofs, are novel to our paper. From a technical perspective, the proof of Theorem \ref{thm-VR-caste} closely follows the proof strategy 
of the main characterization result in \cite{sonmez/yenmez:22}, whereas the proof of Theorem \ref{thm-mOA} builds on abstract research on matroid theory in \cite{gale1968}.

Our paper contributes to  a growing literature on reserve systems that starts with  \cite{hayeyi13}. 
The role of processing order of different types of positions in this framework was first studied by \cite{kominers/sonmez:16}
in a theoretical framework, and subsequently  by   \cite{dur/kominers/pathak/sonmez18} in the context of school choice. 
Subsequent research on reserve systems include \cite{ehlers/hafalir/yenmez/yildirim:14}, \cite{echenique/yenmez:15}, \cite{dur/pathak/sonmez20}, 
 \cite{pathak/sonmez/unver/yenmez:20}, \cite{aygunbo21}, \cite{celebi/flynn:2021, celebi/flynn:2022},
\cite{celebi:2022}, \cite{sonmez/yenmez:22, sonmez/yenmez:22b}, and \cite{Almeer/Dur/Harris/Hauser/Phan/Zhang:24}.  

Finally, in addition to the literature on reserve systems,  
our paper also contributes to a large and growing literature on analysis and design of mechanisms that are deployed in settings in which
issues of social, racial and distributive justice are particularly important. Some of the most related  papers in  this literature includes \cite{abdulson03},  \cite{kesten:06, kesten10},
\cite{erdilergin08}, \cite{Haeringer/Klijn:09},  \cite{kojima:12},  \cite{pycia12}, \cite{andersson/svensson:14}, \cite{kamada/kojima:15},
\cite{Delacretaz/Kominers/Teytelboym:23}, \cite{chenkesten17}, \cite{Andersson2019}, \cite{erdilkumano19}, 
\cite{anderson/ehlers/20},  \cite{EhlersMorrill2020}, \cite{Phan/Tierney/Zhou:21}, \cite{Dur/Morrill/Phan:22}, \cite{reny22}, \cite{dogan/erdil:23}, \cite{Grigoryan:23}, \cite{alvarez/romero-medina:24} and \cite{Imamura/Kawase:24}.

\section{Critical Analysis of Majority and Dissenting Opinions in \textit{Janhit Abhiyan (2022)\/}} \label{sec:critical}

Until recently, in India, VR protections were exclusively available to members of socially and educationally disadvantaged classes who suffered from marginalization and discrimination due to their caste identities. 
This provision, as highlighted in the Dissenting Opinion of Justice Bhat for the judgment \textit{Janhit Abhiyan (2022)\/}, was embedded in the Constitution of India as a reparative and compensatory mechanism 
meant to level the field. This norm changed with the enactment of the 103rd Amendment of the Constitution in 2019. 
Under the Amendment, economically weaker sections of society are granted VR protections for up to 10 percent of positions in government jobs and seats in higher education based on their financial incapacity.  
Controversially, the Amendment maintains the non-overlapping structure of VR protections, excluding socially and educationally disadvantaged classes eligible for existing VR protections, 
namely Scheduled Castes (SCs), Scheduled Tribes (STs), and Other Backward Classes (OBCs), from the new provisions.

Shortly after the enactment of the 103rd Constitutional Amendment in 2019, along with many other petitioners, the two NGOs Janhit Abhiyan and Youth For Equality 
challenged the Amendment at a three-judge Bench of the Supreme Court.\footnote{Details and progression  of the case can be found in the following 
\href{https://www.scobserver.in/cases/youth-for-equality-union-of-india-ews-reservation-case-background/}{\textit{Supreme Court Observer\/} link}, last accessed 05/11/2024.}   
Among their main objections was the exclusion of members of SCs, STs, and OBCs from the scope of the EWS category.
Since the case involved substantial questions around the interpretation of the Constitution, it was referred to a five-judge Constitution Bench of the Supreme Court in August 2020. 

In September 2022, the Constitution Bench announced it would address three main issues to determine whether the Amendment violates the basic structure of the Constitution:\
\begin{enumerate}
\item Can reservations be based solely on economic criteria?
\item Can states offer reservations in private educational institutions without government aid?
\item  Are EWS reservations constitutionally invalid for excluding SCs, STs, and OBCs?
\end{enumerate}

In a landmark judgment,  \textit{Janhit Abhiyan (2022)\/} fundamentally reshaped the understanding of affirmative action in India, as argued by many, 
by affirming the validity of the central law providing 10\% reservation to Economically Weaker Sections (EWS) in November 2022. 
In a split verdict, the five-judge Constitution Bench ruled 3-2 in favor of the 103rd Constitutional Amendment.

The judgment \textit{Janhit Abhiyan (2022)\/} is memorable in many ways. It became the first Supreme Court case in India to have its proceedings live-streamed.\footnote{See \href{https://www.thehindu.com/news/national/supreme-court-live-streams-constitution-bench-proceedings/article65940871.ece}{The Hindu story} ``In a first, Supreme Court live-streams Constitution Bench proceedings,'' last accessed 05/11/2024.} 
The verdict was announced on the last working day of Justice Uday Lalit, who headed the Constitution Bench as the 49th Chief Justice of India (CJI).
According to Supreme Court Observer, a non-partisan platform that reports and analyzes judgments of the Supreme Court, the role of the CJI in the judgment was also unconventional:\footnote{See \href{https://www.scobserver.in/reports/ews-reservation-judgment-sc-upholds-103rd-Amendment-in-3-2-split-verdict/}{Supreme Court Observer article} ``EWS Reservation Judgment: SC Upholds 103rd Amendment in 3-2 Majority,''  last accessed 05/11/2024.}
\smallskip
\begin{quote}
``Chief Justice Lalit’s role in this Judgment, on the eve of his retirement, is unusual for two reasons. First, he has not written an opinion of his own, choosing to express his agreement with Justice Bhat instead. Second, he forms part of the dissent in this case. Chief Justices more often form the Majority Opinion in Constitution Bench cases.''
\end{quote}
But, perhaps, what made the judgment \textit{Janhit Abhiyan (2022)\/} especially memorable is the tone of the
the Dissenting Opinion by Justices Bhat  and Lalit who declared that ``this court has for the first time, in the seven decades of the republic, sanctioned
an avowedly exclusionary and discriminatory principle''.\footnote{See the 
\href{https://www.scobserver.in/wp-content/uploads/2021/10/EWS-Reservations-Judgment-299-399.pdf}{opening paragraph of the Dissenting Opinion}.}  

In this section, we discuss both the Majority Opinion and the Dissenting Opinion in this historical case, relating them to our formal analysis in Section \ref{sec:3choicerules}. 
Among other findings, we show that several key arguments used in the Majority Opinion are outright false. 
Since the majority judgment on the third issue considered by the Bench—whether SC/ST/OBC can be excluded from the scope of EWS—is directly linked to these inaccurate arguments, 
we believe these mistakes played a pivotal role in this historically significant judgment. Furthermore, we argue that our analysis in Section \ref{sec:3choicerules} 
can offer valuable insights for formulating a less divisive resolution for the crisis in the future.

\subsection{The Disagreement Between the Majority and Dissenting Opinions}

All five justices declared that reservations can be granted solely on the basis of economic criteria. Their disagreement was mainly on the constitutionality of the exclusion of 
SC, ST, and OBC from the scope of the EWS reservation. The two dissenting justices declared the exclusion of socially and educationally disadvantaged classes 
from the scope of EWS reservation to be unconstitutional. In contrast, the three majority justices declared it to be constitutional, arguing that the exclusion 
was necessary to provide positive discrimination to groups not covered by earlier reservations.

The two minority justices characterized the exclusion as ``Orwellian'' and expressed their strong objection in the following concluding paragraphs of their Dissenting Opinion:\smallskip
\begin{indquote}
``191. A universally acknowledged truth is that reservations have been
conceived and quotas created, through provision in the Constitution, only
to offset fundamental, deep rooted generations of wrongs perpetrated on
entire communities and castes. Reservation is designed as a powerful tool 
to enable equal access and equal opportunity. Introducing the economic
basis for reservation - as a new criterion, is permissible. Yet, the ``othering''
of socially and educationally disadvantaged classes -
including SCs/ STs/OBCs by excluding them from this new reservation on the ground that they
enjoy pre-existing benefits, is to heap fresh injustice based on past
disability. [...] The net effect of the entire exclusionary
principle is Orwellian, (so to say) which is that all the poorest are entitled
to be considered, regardless of their caste or class, yet only those who
belong to forward classes or castes, would be considered, and those from
socially disadvantaged classes for SC/STs would be ineligible.  [...] \smallskip

\noindent192. Therefore, on question 3, it is clear that the impugned Amendment
and the classification it creates, is arbitrary, and results in hostile
discrimination of the poorest sections of the society that are socially and
educationally backward, and/or subjected to caste discrimination. For
these reasons, the insertion of Article 15(6) and 16(6) is struck down, is
held to be violative of the equality code, particularly the principle of nondiscrimination and non-exclusion which forms an inextricable part of the
basic structure of the Constitution.''
\end{indquote}

\subsection{Alleged Necessity of the Exclusion to Avoid ``Double Benefits''} \label{sec:doublebenefits}
  
The majority justices were not unsympathetic to the perspective provided by the two dissenting justices above. 
In paragraph 77 of the Majority Opinion, Justice Maheshwari articulated their perspective as follows.:\smallskip

\begin{indquote}
``77.1. [...] Rather, according to the petitioners,
the classes covered by Articles 15(4), 15(5) and 16(4) are comprising of
the poorest of the poor and hence, keeping them out of the benefit of
EWS reservation is an exercise conceptionally at conflict with the
constitutional norms and principles.\smallskip

\noindent 77.2. At the first blush, the arguments made in this regard appear to be
having some substance because it cannot be denied that the classes
covered by Articles 15(4), 15(5) and 16(4) would also be comprising of
poor persons within. However, a little pause and a closer look makes it
clear that the grievance of the petitioners because of this exclusion
remains entirely untenable and the challenge to the Amendment in
question remains wholly unsustainable. As noticed infra, there is a
definite logic in this exclusion; rather, this exclusion is inevitable for the
true operation and effect of the scheme of EWS reservation.''
\end{indquote}

Therefore, while agreeing that the exclusion may not correspond to an ideal scenario, the majority justices hold the (factually inaccurate) opinion 
that it is inevitable from an operational perspective. In paragraphs 79-82 of their Majority Opinion, Justice Maheshwari articulates the alleged necessity of the exclusion as follows:\smallskip

\begin{indquote}
``79. [...] The moment there is a vertical reservation, exclusion is the
vital requisite to provide benefit to the target group. In fact, the affirmative
action of reservation for a particular target group, to achieve its desired
results, has to be carved out by exclusion of others. [...] But for this exclusion, the purported
affirmative action for a particular class or group would be congenitally
deformative and shall fail at its inception. Therefore, the claim of any
particular class or section against its exclusion from the affirmative action
of reservation in favour of EWS has to be rejected.\smallskip

\noindent 80. [...] It could easily be seen
that but for this exclusion, the entire balance of the general principles of
equality and compensatory discrimination would be disturbed, with extra
or excessive advantage being given to the classes already availing the
benefit under Articles 15(4), 15(5) and 16(4).\smallskip

\noindent 81. Putting it in other words, the classes who are already the recipient
of, and beneficiary of, compensatory discrimination by virtue of Articles
15(4), 15(5) and 16(4), cannot justifiably raise the grievance that in
another set of compensatory discrimination for another class, they have
been excluded. It gets, perforce, reiterated that the compensatory
discrimination, by its very nature, would be structured as exclusionary in
order to achieve its objectives. Rather, if the classes for whom affirmative
action is already in place are not excluded, the present exercise itself
would be of unjustified discrimination.\smallskip

\noindent 82.1. [...] As said above, compensatory
discrimination, wherever applied, is exclusionary in character and could
acquire its worth and substance only by way of exclusion of others. Such
differentiation cannot be said to be legally impermissible; rather it is
inevitable. When that be so, clamour against exclusion in the present
matters could only be rejected as baseless.''\smallskip
\end{indquote}

Observe that the sole justification given by the majority justices for their support of the exclusion of the socially and educationally disadvantaged classes from the scope of 
EWS is a flawed technical argument made in paragraphs 79-82 of the Majority Opinion.\footnote{This observation—i.e., the justification offered by Justice Maheshwari 
in support of the exclusion—is also acknowledged by the analysis in the Supreme Court Observer report  \href{https://www.scobserver.in/reports/ews-reservations-judgment-matrix/}{``EWS Reservations: Judgment Matrix,''} 
last accessed on 05/11/2024.}
They erroneously argue that the exclusion of beneficiaries of earlier provisions from a new provision is absolutely necessary to deliver benefits to individuals outside the scope of earlier provisions. 
Additionally, they argue that the inclusion of the socially and educationally disadvantaged classes in the scope of EWS as well would result in excessive advantage for members of these groups. 
This latter argument is also referred to as the ``double-benefit argument'' by the justices, government officials, and media.\footnote{See, for example,
\href{https://www.livelaw.in/top-stories/scstobc-exclusion-from-ews-quota-logical-necessary-to-avoid-double-benefits-supreme-court-213544}{Live Law story} 
``SC/ST/OBC Exclusion From EWS Quota Logical, Necessary To Avoid Double Benefits: Supreme Court,'' last accessed 05/11/2024.} 

While the technical justification offered by the majority justices is accurate under non-overlapping VR protections, it is false under overlapping VR protections, 
i.e., the relevant version of the problem with the scope-extended EWS category. Since one of the main issues the bench was deciding was the constitutionality of the exclusion, 
from a formal perspective, it was also deciding the constitutionality of the non-overlapping structure of VR protections. 
Despite this, in their decision, the majority justices relied on arguments that critically depend on an underlying environment with non-overlapping VR protections. 
In a manner of speaking, the majority justices proved the necessity of an assumption by using the assumption itself.

Specifically, as demonstrated in Theorems  \ref{thm-EWS} and \ref{thm-EWS2}, EWS positions are predominantly allocated to individuals not covered by earlier caste-based VR protections under the EWS-last VR-processing policy, 
except when doing so would violate the Equality Principle. Furthermore, as indicated in  Theorem \ref{thm-EWS2}, under this policy, members of socially and educationally disadvantaged classes 
benefit from an additional EWS membership only if their membership in caste-based VR-protected categories puts them at a disadvantage compared to membership in EWS. 
This is why we describe the EWS-last VR processing policy as one that effectively provides conditional EWS membership to beneficiaries of caste-based VR protections in Section \ref{sec:conditional}. 
Consequently, the technical justification provided by the majority justices for their controversial decision stems entirely from their oversight of the implications of overlapping VR protections, 
a technical and subtle phenomenon with which the justices are naturally not familiar.

It is important to highlight two points at this juncture. First of all, as presented in Section \ref{sec:EWS-first}, the scope-extended EWS policy indeed provides 
additional benefits for the socially and educationally disadvantaged classes under the alternative EWS-first VR processing policy. 
Thus, we are not suggesting that the technical point made by the majority justices is never valid, regardless of how EWS positions are processed. 
However, we do contend that their argument is entirely false under the EWS-last VR processing policy. Therefore, in our view, the core of the arguments made by 
Justice Maheshwari in support of the exclusion of SC/ST/OBC from the scope of EWS is more consistent with the removal of the exclusion, albeit under the EWS-last VR processing policy. 
Our view here is also consistent with our formal results in Theorems \ref{thm-EWS} and \ref{thm-EWS2}, which establishes that the EWS-last VR 
processing policy is the minimal possible deviation from the existing system that avoids violating the Equality Code.

Secondly, we are not suggesting that the dissenting justices would necessarily be in favor of the EWS-last VR processing policy 
if the majority justices had supported this policy due to the advantage it provides to individuals who do not benefit from earlier provisions. 
On the contrary, as argued in Section \ref{sec:EWS-first}, we believe the Dissenting Opinion is more aligned with the EWS-first VR processing policy. 
Our belief is based on the following three arguments in the Dissenting Opinion. In paragraph 100 of his Dissenting Opinion, Justice Bhat rejects ``the double benefit argument'' 
for a reason completely different from our technical objection:\smallskip  
\begin{indquote}
``100. The characterisation of including the poor (i.e., those who qualify
for the economic eligibility) among those covered under Articles 15(4) and
16(4), in the new reservations under Articles 15(6) and 16(6), as bestowing
`double benefit' is incorrect. What is described as `benefits' for those
covered under Articles 15(4) and 16(4) by the Union, cannot be understood
to be a free pass, but as a reparative and compensatory mechanism meant
to level the field - where they are unequal due to their social stigmatisation.
This exclusion violates the non-discrimination and the non-exclusionary
facet of the equality code, which thereby violates the basic structure of the
Constitution.'' \smallskip
\end{indquote}

Here, Justice Bhat argues that the earlier provisions do not disqualify members of socially and educationally disadvantaged classes from deprivation-based 
provisions because the earlier provisions are reparative and compensatory. He further elaborates on this point in paragraph 168 of his Dissenting Opinion.\smallskip

\begin{indquote}
``168. The characterisation of reservations for economically weaker
sections of the population (EWS) as compensatory and on par with the
existing reservations under Articles 15(4) and 16(4), in my respectful
opinion, is without basis.'' \smallskip
\end{indquote}

This latter argument by Justice Bhat suggests that he sees the earlier reparative and compensatory reservations as a ``higher-level'' provision, 
which is consistent with the EWS-first VR processing policy. 

Finally, as discussed earlier in Section \ref{sec:EWS-first}, in paragraph 191 of his Dissenting Opinion, he makes points that motivated 
our axiom of ``respect for mobility from reparative categories to EWS,'' as given in Definition \ref{def-VR-caste}.\smallskip
\begin{indquote}
``Thirdly, it denies the chance of mobility from the reserved quota (based on past discrimination) to a reservation benefit based
only on economic deprivation.'' \smallskip
\end{indquote}

In light of these arguments, the Dissenting Opinion is more aligned with the expanded-scope EWS category under the EWS-first VR processing policy. 

\subsection{Interpretation of the Principle of Equality and the Cut-off Crisis}

Another disagreement between the Majority Opinion and the Dissenting Opinion pertains to the interpretation of the \textit{Principle of Equality\/}. 
Majority Justice Maheshwari explains this important principle and how it can be constitutionally applied, referred to as \textit{reasonable classification}, as follows:\smallskip
\begin{indquote}
``44. In a nutshell, the principle of equality can be stated thus: equals
must be treated equally while unequals need to be treated differently,
inasmuch as for the application of this principle in real life, we have to
differentiate between those who being equal, are grouped together, and
those who being different, are left out from the group. This is expressed
as reasonable classification. Now, a classification to be valid must
necessarily satisfy two tests: first, the distinguishing rationale should be
based on a just objective and secondly, the choice of differentiating one
set of persons from another should have a reasonable nexus to the object sought to be achieved.''\smallskip
\end{indquote}

Further elaborating on the Principle of Equality is Majority Justice Bela Trivedi:\smallskip
 \begin{indquote}
``20. [...]  Treating economically weaker sections of the citizens as a separate class would be a reasonable
classification, and could not be termed as an unreasonable or unjustifiable
classification, much less a betrayal of basic feature or violative of Article 14.
As laid down by this Court, just as equals cannot be treated unequally,
unequals also cannot be treated equally. Treating unequals as equals would as
well offend the doctrine of equality enshrined in Articles 14 and 16 of the Constitution.''\smallskip
\end{indquote}

According to majority justices, individuals facing economic deprivation is a reasonable classification as a class. 
This point is agreed upon by the dissenting justices as well. 
The majority justices, however, further refine this class by excluding the socially and educationally disadvantaged classes. 
This is where the fundamental difference lies between the majority justices and dissenting justices. 
In Section \ref{sec:doublebenefits}, we presented the main, albeit flawed, technical justification of this exclusion from the perspective of the majority justices.
A second justification is offered by the majority justices based on the principle of 
justice, where they argue that ``treating unequals as equals would as well offend the doctrine of equality.'' 
According to this argument, individuals from the socially and educationally disadvantaged classes are different from 
individuals from the upper classes, therefore economically deprived members of the two groups are also different, 
and consequently the two groups have to be treated differently. But what does the phrase ``unequals cannot be treated equally'' exactly mean in this context?
Does it allow for treating disadvantaged groups worse than advantaged groups? 
We are fairly confident that this is not what is intended in the constitution by the principle of equity. 
Clearly ``unequals cannot be treated equally''  must mean that disadvantaged individuals should be treated better.
This interpretation is the basis of our axiom of ``compliance with the Equality Principle'' as formulated in Definition \ref{axiom:equality-code}.  

If our interpretation is accurate, then a systematic violation of this principle is enabled by the exclusion of socially and educationally disadvantaged classes from EWS reservations.
After all, the Amendment allows for an economically deprived  member of a privileged forward class to receive a position with a low merit score, 
while it denies the same position for an economically deprived member of a disadvantaged class who has a higher merit score.

According to Theorem \ref{thm-EWS}, the scope-increased EWS with EWS-last VR processing policy represents the smallest deviation from the current Amendment that avoids this anomaly. 
This violation of the Principle of Equality is not merely theoretical but a regular occurrence under the Constitutional Amendment, as highlighted in the story ``EWS verdict shows merit matters only when it’s ‘their’ children, not ‘our’ kids'' 
from the digital platform  \href{https://theprint.in/opinion/ews-verdict-shows-merit-matters-only-when-its-their-children-not-our-kids/1206345/}{\textit{The Print\/}}. 
Indeed, as it is also highlighted as follows in this story, this anomaly is a recurring phenomenon under the contested Amendment: \smallskip

\begin{quote}
``Finally, the judgment makes no mention of the most frequently cited principle when discussing reservation, `merit'. 
This silence is stark in the face of the evidence of the first two years of application of EWS quota that the cut-off for EWS was even lower than the OBC. 
Clearly merit matters only when we discuss `their' children, not when we discuss `our' children
 benefiting from capitation fees, securing spurious degrees abroad or obtaining EWS quota.''
 \end{quote}
 
That is,  for the first two years of the implementation of EWS reservation, the minimum score needed to secure a position 
was lower for the economically deprived members of forward classes compared to the economically deprived members of 
the disadvantaged classes. 
Indeed, this anomaly is one of the three reasons for the Tamil Nadu government's opposition to EWS reservation.\footnote{See the
\href{https://www.thenewsminute.com/article/three-reasons-why-dmk-has-opposed-10-ews-reservation-167786}{\textit{TNM\/} story}
``Three reasons why the DMK has opposed 10\% EWS reservation'', last accessed 05/11/2024.\label{footnoteTN}}

\section{Epilogue: Informed Neutrality in Minimalist Market Design} \label{sec:conclusion}

India's affirmative action system, constitutionally protected, is characterized by a complex set of normative goals and requirements. 
Historically, justices at the Supreme Court and state high courts have excelled in formulating these principles and guiding their implementation. 
However, due to the sheer complexity of the issue, they have sometimes failed to fully grasp the collective implications of these principles or understand how 
changes in various aspects of their application might interfere with them. This has led to unintended consequences, 
including major disputes between groups, inconsistencies between judgments, loopholes in the system, and large-scale disruptions in recruitment and school admissions \citep{sonmez/yenmez:22, sonmez/yenmez:22b}.

In this paper, we analyze one of the most significant episodes of these unintended consequences. In a 3-2 split verdict, the Supreme Court 
fundamentally altered the nature of affirmative action in India through its judgment in \textit{Janhit Abhiyan (2022)\/}. 
Reflecting the controversy, Justice Bhat, in his dissenting opinion, stated that ``the impugned amendment and the classification it creates, 
is arbitrary, and results in hostile discrimination against the poorest sections of society that are socially and educationally backward, and/or subjected to caste discrimination.''

What is particularly striking about the judgment is that the majority justices used purely technical arguments to justify their controversial decision on the constitutionality 
of excluding socially and educationally backward classes from the scope of EWS. One of our contributions in this paper is revealing that these arguments are 
irrefutably false.\footnote{Confusion among policymakers is particularly common in real-life applications of reserve systems. 
See, for example, \cite{dur/kominers/pathak/sonmez18}, \cite{pathak/rees-jones/sonmez:20, pathak/rees-jones/sonmez:23}, and \cite{sonmez/yenmez:22}.} 
The irony of this situation aligns with a crucial insight from  \cite{Li2017}: Policymakers may understand the details of their environment yet struggle to articulate their ethical requirements precisely. 
In the case of \textit{Janhit Abhiyan (2022)\/}, the majority justices failed to assess the implications of a significant change in the fair allocation problem they were tasked to resolve, 
specifically the transition of VR-protected categories from non-overlapping to overlapping.

As further emphasized in \cite{Li2017}, economists, with their expertise in formal modeling and analysis, have a comparative advantage in identifying the normatively relevant properties of complex systems. 
Utilizing this advantage, we demonstrate that this crisis could have been resolved much more amicably with at least three alternative policies, 
each avoiding the failures of the existing system and backed by a distinct normative justification. 
Our paper underscores that expertise in formal modeling and analysis is not only valuable but also necessary for making informed societal decisions involving complex systems.

In our analysis, we utilized \textit{minimalist market design} \citep{sonmez:23}, an institution design paradigm developed for applications where researchers can inform and influence policy as outsider critics, 
often in settings where decision-makers may have vested interests in maintaining the status quo. Evolved through experiences in school choice and kidney exchange, 
this paradigm has been successfully deployed in recent years in a wide range of applications such as the US Army's reform of its branching system in 2020 \citep{greenberg/pathak/sonmez:24}, 
the fair allocation of vaccines and therapies during the COVID-19 pandemic \citep{pathak/sonmez/unver/yenmez:20}, 
and the deployment of the world's largest liver exchange system in Turkey in 2022 \citep{yilmaz/sonmez/unver:2023, yilmaz/sonmez/unver:2024}.

The key to the success of this paradigm lies in the ``persuasion strategy'' embedded within its three main tasks. 
Rather than merely relying on mainstream principles in economics (e.g., preference utilitarianism), the first task involves identifying the objectives of various stakeholders 
(e.g., policymakers, decision-makers), which we refer to as the institution's \textit{mission\/}. The second task is assessing whether the existing institution meets its underlying mission, 
and if not, identifying the \textit{root causes\/} of any discrepancies between what is desired and what is achieved. 
The third task involves designing a new institution—often by developing ``custom theory''—that fulfills the mission by addressing only these root causes and nothing else, 
hence the term ``minimalist'' as the defining feature of this paradigm. This approach has often been crucial in developing trust between various stakeholders and design economists who act as outsider critics.

While typically not central to persuasion, our case study highlights the importance of an additional task in applications with potentially multiple minimalist interventions: 
Maintaining \textit{informed neutrality\/}  \citep{Li2017} between reasonable normative principles served by alternative designs, in addition to the primary mission of 
the underlying institution.\footnote{The term \textit{informed neutrality} is coined in \cite{Li2017}  in the context of ethical principles, who advocated for its broader practice in market design.} 
Moreover, in some applications, strictly adhering to this approach may be a ``moral necessity'' for market designers who wish to participate in these crucial decisions 
without introducing various forms of ``algorithmic bias'' into the underlying institution. Such biases could systematically benefit certain groups at the expense of others \citep{vanBasshuysen:2022}. 
Additionally, failing to adhere to this approach may result in a ``normative gap''—a divergence between the objectives of policymakers and the constructs of economic engineers \citep{hitzig2018, hitzig2020}.

\bibliographystyle{econ-aea}
\bibliography{Overlapping-VR-v15-ArXiv} \bigskip

\newpage

\begin{appendix}
\begin{center}
{\large \textbf{Appendices}}
	
\end{center}

\section{Technical Preliminaries for the Proofs} 
\label{app:preliminaries}

\subsection{Matchings, Matroids, Greedy Algorithm, and Choice Rules} \label{app:preliminaries-1}

In this subsection, we introduce auxiliary mathematical structures that will be useful in formalizing some of our definitions and proving our results.

Fix a profile of category memberships $\rho \in \mem$. Let $\cale^c=\cale^c(\rho)$ for each $c \in \calr$.

Let $\calq$ be the set of reservation vectors.  Consider a reservation vector $q=\big (q^\nu \big )_{\nu \in \calv} \in \calq$. 

\begin{definition} \label{defn:matching} Let $I \subseteq \cali$. 	A \textbf{matching} is a mapping $\mu : I \rightarrow \calv  \medcup \big \{\emptyset \big \}$ such that 
	\begin{itemize}
		\item for each $i \in I$, $\mu(i)=\emptyset$ \; or \; $\mu(i)=v$ with $i \in \cale^v $, and
		\item for each $v \in \calv$, $|\mu^{-1}(v)| \le {q}^{v}$.
	\end{itemize}
	For an individual $i$, $\mu(i)=\emptyset$ refers to her remaining \textbf{unmatched}. Let $\bbm(q,I)$ be the set of matchings for $I$ under $q$.\footnote{Although we do not change the real reservation vector $q$ after it is set,  we keep a reservation vector in the argument of the set of matchings. This is because we introduce auxiliary reservation vectors by modifying $q$, which play an important role in formal definitions and proofs.} 
\end{definition}
The set of individuals matched by a matching $\mu \in \bbm(q,I)$ with a category $v \in \calv$ is denoted as 
\[
	I_{\mu,v}=\mu^{-1}(v)=\Big \{ i \in I : \mu(i) = v   \Big \},  
\]
and the overall set of individuals matched by 
$\mu$ is denoted as
\[
 I_{\mu}=\mu^{-1}(\calv)=\bigcup_{\nu \in \calv} I_{\mu,\nu}.
\]

Using the auxiliary concept of matchings, we define the following  function, which was formally defined for the special case of non-overlapping vertical reservations before:

\begin{definition} \label{defn:VR-max} The \textbf{VR-maximality function} ${\beta: 2^\cali \rightarrow \mathbb{N}}$ is defined for any $I \subseteq \cali$ as
	\[
	 	\beta(I)=\max_{\mu \in \bbm(q,I)} \Big | \big \{i\in I : \mu(i)\in \calr  \big \}\Big |. 
	\]
\end{definition} 
$\beta(I)$ is the maximum number of individuals in $I$ that can be accommodated by VR-protected positions.

The following mathematical structure helps us understand the properties of our choice rules.

\begin{definition} A pair of finite sets $(E, \calx)$ is a \textbf{matroid} if $\calx \subseteq 2^E$ and 
\begin{enumerate}
	\item $\emptyset \in \calx$,
	\item if $F \in \calx$, then for any $G \subsetneq F$ we have $G \in \calx$, and 
	\item if $F,G \in \calx$ such that $|G|<|F|$, then there is an element $x \in F \setminus G$ such that $G \cup \{x\} \in \calx$.   
\end{enumerate}
Each element of $\calx$ is called an \textbf{independent set} of matroid $(E,\calx)$. 
\end{definition}

Next, we define a specific matroid for our purposes.

For any set of individuals $I\subseteq \cali$, define a collection of subsets of $I$
\begin{align}
	 \boldsymbol{\cala(q,I)}=\big \{J \subseteq I : \; \exists\, \mu \in \bbm(q,I) \; \text{s.t.} \; J \subseteq I_{\mu} \big \}.
\end{align} 
We refer to each set $J \in \cala(q, I)$ as an  \textbf{assignable subset of $\boldsymbol{I}$ under $\boldsymbol{q}$}.

\begin{lemma}[\citet{edmonds/fulkerson:65}] \label{lem-transversal} 	For any set of individuals $I\subseteq \cali$, the pair $\big(I,\cala(q,I)\big )$ is a matroid.
\end{lemma}
Matroid $\big (I,\cala(q,I) \big )$ is called a \textbf{transversal matroid} \citep[e.g., see][]{lawler}. Observe that assignable subsets of $I$ under $q$ are exactly the independent sets of transversal matroid $\big (I,\cala(q, I) \big )$.
 
Let $\scores \subsetneq \mathbb{R}_+^{\big|\cali\big|}$ be the set of merit score vectors, which are non-negative real-valued, and no two individuals have the same score.   Given a merit score vector $\sigma \in \scores$,
we find a particular assignable subset of $I$ under $q$ that we denote as ${G(q,I;\sigma)}$ through an iterative procedure.\footnote{The greedy algorithm is defined for general matroids, although we define it using special wording for a transversal matroid. Lemma \ref{lem-Gale} also holds for the greedy algorithm for general matroids.} 

\begin{quote}
	\noindent{}\textbf{Greedy algorithm for  transversal matroid $\big (I,\cala(q,I) \big )$ under merit score vector $\sigma$:} 

 	\noindent{}\textbf{Step 0 (Initiation):} Let $I_0=\emptyset$. 
 
 	\noindent{}\textbf{Step k $\big(\boldsymbol{k \in \{1,\hdots,|I|+1\}\big)}$:} Assuming such an individual exists, choose the highest $\sigma$-score individual whom we denote as $i_k \in   I \setminus I_{k-1} $	such that $I_{k-1}\cup \{i_k\} \in \cala(q,I)$ and let $I_k= I_{k-1}\cup \{i_k\}$. If no such individual exists, then end the process by setting $\boldsymbol{G(q,I;\sigma)}=I_{k-1}$.		
\end{quote}
We refer to the outcome set $G(q,I;\sigma)$ of this procedure as \textbf{the greedily assignable subset in $\boldsymbol{\cala(q,I)}$ under $\boldsymbol{\sigma}$}.

While the greedily assignable subset is uniquely defined, there can be multiple matchings that assign the same set of individuals to different categories. We refer to any matching $\mu \in \bbm(q,I)$ such that
	$$
		I_{\mu} = G(q,I;\sigma)
	$$
as \textbf{a greedy matching in $\boldsymbol{\bbm(q,I)}$ under $\boldsymbol{\sigma}$}.
The following important property of the greedy assignable subset is critical in proving Theorem \ref{thm-mOA}.

\begin{lemma}[\citet{gale1968}]  \label{lem-Gale} Consider reservation vector $q \in \calq$, merit score vector $\sigma \in \scores$, set of individuals $I \subseteq \cali$, and assignable subset $J \in \cala(q,I)$, the greedily assignable subset  $G(q,I;\sigma)$ Gale dominates assignable subset $J$. 
\end{lemma} 
An immediate corollary to this result is as follows using the definition of Gale domination.
\begin{corollary}\label{cor-greedy-max}
	For any reservation vector $q \in \calq$, merit score vector $\sigma \in \scores$, set of individuals $I \subseteq \cali$, and assignable subset $J \in \cala(q,I)$, the greedily assignable subset  $G(q,I;\sigma)$ satisfies
	$\big|G(q,I;\sigma)\big|\ge |J|$.
\end{corollary}

\begin{remark}\label{rem-mOA}
	The meritorious O\&A choice rule $C_{\o}(\rho;.)$ utilizes the greedy algorithm for any $I \subseteq \cali$  for transversal matroid $\big (J,\cala({q'},J)\big )$  under $\sigma$ in Step 2 of  its procedure where 
	\begin{itemize}
		\item $J=I \setminus C_{\o}^o(\rho;I)$, and
		\item $q' \in Q$ is obtained from $q$ by setting (i) ${q'}^o=0$ and (ii) ${q'}^c=q^c$ for each $c \in\calr$. 
	\end{itemize} 		
	We pick a greedy matching  $\mu' \in \bbm(q',J)$ under $\sigma$ to determine the VR-protected category components of $C_{\o}(\rho;.)$. For each $c \in \calr$, we set
		$$
			C_{\o}^c(\rho;I)=I_{\mu',c}. 
		$$ 
	Thus, 
		$$\bigcup_{c \in\calr} C_{\o}^c(\rho;I)=G(q',J;\sigma).
		$$
\end{remark} 

\subsection{A Property of Choice Rules}

We explicitly state an implication of Proposition \ref{prop0} for single-category choice rules. 
This result is implied by \citet{sonmez/yenmez:22} and is built on a more layered setup that also includes horizontal reservations used in India. 
However, we prove it here with a simpler proof for self-containment of the paper with our less layered setup: 
\begin{lemma2} \label{basic-lem0}
	For any $\rho \in \mem$ and
 $v \in \calv$, a single-category choice rule $C^v (\rho;.)$ is non-wasteful and satisfies no justified envy if and only if $C^v(\rho;.)= C_{sd}^v(\rho;.)$.
\end{lemma2}
\begin{proof}[Proof of Lemma \ref{basic-lem0}]
Let $\rho \in \mem$ and
 $v \in \calv$. If $v \in \calr$, let $\cale^v=\cale^v(\rho)$.\smallskip 
 
 First, observe that $C_{sd}^v(\rho;.)$ is non-wasteful, as in the procedure for a set of agents $I \subseteq \cali$, it chooses  $\min\Big \{\big |I \cap \cale^v\big |, \; q^v \Big \}$ individuals in $I \cap \cale^v$. It also satisfies no justified envy, as each individual it chooses from a set of agents $I \subseteq \cali$, i.e., an individual in $C_{sd}^v(\rho;I)$, has a higher merit score than any individual in  $\big (I \medcap \cale^v\big ) \setminus C_{sd}^v(\rho;I)$. \smallskip 
 
 Conversely, suppose $C^v (\rho;.)$ satisfies non-wastefulness and no justified envy. Toward a contradiction, suppose there exists some $I \subseteq \cali$ such that $C^v (\rho;I) \neq C_{sd}^v(\rho;I).$   
 
 First, we establish $\big|C^v (\rho;I)\big| \ge \big|C_{sd}^v(\rho;I)\big|$. On the contrary, if $\big|C^v (\rho;I)\big|<\big|C_{sd}^v(\rho;I)\big|$ then there exists some individual $j \in C_{sd}^v(\rho;I) \setminus C^v (\rho;I) \subseteq I \medcap \cale^v$. 
 Since $\big|C_{sd}^v(\rho;I)\big|\le q^v$, $\big|C^v (\rho;I)\big|<q^v$. This contradicts non-wastefulness of $C^v(\rho;.)$. Thus, $|C^v (\rho;I)|\ge |C_{sd}^v(\rho;I)|$. 
 
 Similarly, we show that $\big |C^v (\rho;I)\big | \le  \big|C_{sd}^v(\rho;I)\big|$. These two imply $\big|C^v (\rho;I)\big|= \big|C_{sd}^v(\rho;I)\big|$.  
  
 Hence, there exists some $j \in C^v (\rho;I) \setminus C_{sd}^v(\rho;I)$ and $j' \in C_{sd}^v(\rho;I) \setminus  C^v (\rho;I)$. If $\sigma_j > \sigma_{j'}$ then this contradicts no justified envy property of $C_{sd}^v(\rho;.)$ and if $\sigma_j < \sigma_{j'}$ then this contradicts no justified envy property of $C^v(\rho;.)$. 
 Thus, such $j,j'$ cannot exist and hence,   $C^v (\rho;I) = C_{sd}^v(\rho;I)$, completing the proof.
\end{proof}

On the other hand, the following fundamental lemma is used in proving several of our results:

\begin{lemma} \label{basic-lem0-2}
	Fix $\rho \in \mem$. If $C(\rho;.)=(C^\nu(\rho;.))_{\nu \in \calv}$ is a multi-category choice rule that satisfies  non-wastefulness and no justified envy, then for any $I \subseteq \cali$ and $v \in \calv$,
	$$ C^v(\rho;I)= C_{sd}^v \Big (\rho; I\setminus \bigcup _{\nu \in \calv\setminus \{v\}} C^\nu(\rho;I) \Big).$$
\end{lemma}

\begin{proof*}[Proof of Lemma \ref{basic-lem0-2}]
	Fix $\rho \in \mem$. Let $C(\rho;.)$ be a choice rule that satisfies non-wastefulness and no justified envy. Let $I \subseteq \cali$ and $v \in \calv$. Let $\cale^v=\cale^v(\rho)$ if $v \in \calr$. 
	Let $J=I\setminus \bigcup _{\nu \in \calv\setminus \{v\}} C^\nu(\rho;I) \Big)$.

	First, suppose there exists $i \in C^{v}(\rho;I) \setminus C_{sd}^v(\rho;J)$. Then there exists some $j \in C_{sd}^v(\rho;J) \setminus C^{v}(\rho;I)$, as otherwise $|C_{sd}^v(\rho;J)|<q^v$ and $i \in \big (J \medcap \cale^v \big ) \setminus C_{sd}^v(\rho;J)$ contradict non-wastefulness of $C_{sd}^v(\rho;.)$ established in Lemma \ref{basic-lem0}. Then by no justified envy property of $C_{sd}^v(\rho;.)$ established in Lemma \ref {basic-lem0}, $\sigma_j > \sigma_i$. Since $j \in J \cap \cale^v$ and $j \notin C^{v}(\rho;I)$, then $j \in \big  (I \medcap \cale^v \big ) \setminus \h{C}(\rho;I)$. This contradicts no justified envy property of $C(\rho;.)$. Thus, such an agent $i$ does not exist.
	
	Next, suppose there exists $i \in C_{sd}^v(\rho;J) \setminus C^{v}(\rho;I)$. Then there exists some $j \in C^{v}(\rho;I) \setminus C_{sd}^v(\rho;J)$, as otherwise $\big |C^{v}(\rho;I)\big |<q^v$, $i \in (J \medcap \cale^v)$, and $i \notin \h{C}(\rho;I)$ contradict non-wastefulness of $C(\rho;.)$. We also have $i \in (I \medcap \cale^v)\setminus \h{C}(\rho;I)$. Then  $\sigma_j > \sigma_i$ by no justified envy property of $C(\rho;.)$.  This contradicts no justified envy property of $C_{sd}^v(\rho;.)$ established in Lemma \ref{basic-lem0}. Thus, such an agent $i$ does not exist.
	
	This concludes the proof of  	$ C^v(\rho;I)= C_{sd}^v(\rho; J)$ and the lemma.
\end{proof*}

\section{Proofs} \label{app:proofs}

\subsection{Proofs of Preliminary Results in Section \ref{sec:model}} $~$\\ 

\begin{proof*}[Proof of Lemma \ref{basic-lem1}] 
	Fix  $\rho \in \mem$ and $\rhd\in\Delta$.  Let $\cale^c=\cale^c(\rho)$ for each $c\in \calr$. 	Fix also $I\subseteq\cali$. \smallskip
	
	\textit{Non-wastefulness}: Suppose $v \in \calv$ be such that  $|C_\rhd^{v} (\rho;I)|<q^{v}$,
	and there exists some $j\in I\setminus\h{C}_\rhd (\rho;I)$.
	Then, just before $v$ is processed in the sequence $\rhd$,
	$j$ is still available. Moreover, she does not receive a category-$v$ position. Thus, even though  the serial dictatorship outcome for category $v$, $C_{sd}^{v}\Big (\rho; I \setminus \bigcup_{\nu \in \calv : \nu \mathrel{\lhd} v} C_\rhd^{\nu} (\rho;I) \big ), $ leaves some vacant positions in category
	$v$, $j$ does not receive a position. Thus, $j\notin\cale^v$ by the non-wastefulness of serial dictatorship through Lemma \ref{basic-lem0}. Since by definition 
	\begin{align}
		C_\rhd^{v} (\rho;I)=C_{sd}^{v} \Big(\rho; I \setminus \bigcup_{\nu \in \calv : \nu \mathrel{\lhd} v} C_\rhd^{\nu}(\rho;I) \Big ) \label{eq:lem1}
	\end{align}
 and the argument in the previous sentence is true for each category $v$, the sequential choice rule is non-wasteful. \smallskip

	\textit{No justified envy:} Suppose $v \in \calv$, 
	$i\in C_\rhd^{v} (\rho;I)$, and $j\in\big(I\medcap\cale^{v}\big)\setminus\h{C}_\rhd (\rho;I)$ such that $\sigma_j > \sigma_i$.  Thus, both individuals $i$ and $j$ are available just before category $v$ is processed in the order of precedence $\rhd$. However, as $C_{sd}^v(\rho;.)$ satisfies no justified envy by Lemma \ref{basic-lem0}, then $j$ should have been in $ C_\rhd^{v} (\rho;I)=C_{sd}^v \Big( \rho; I \setminus \medcup_{\nu \in \calv : \nu \mathrel{\lhd} v} C_\rhd^{\nu} (\rho;I) \Big)$ instead of $i$, a contradiction. Thus, the sequential choice rule satisfies no justified envy. 
\end{proof*}

\begin{proof*}[Proof of Lemma \ref{basic-lem2}] 
	Fix $\rho \in \mem$ and $\rhd\in\Delta^{o}$.   
	Suppose  
	$I\subseteq\cali$, $c\in\calr$, and $i\in C_\rhd^{c} (\rho;I)$.
	In executing the procedure of the sequential choice rule according to $\rhd$,  the open category is processed first, and individual $i$ is still available when the VR-protected category $c$ is about to be processed. Thus, it should be the case that   $\big|C_\rhd^{o} (\rho;I)\big|=q^{o}$, as otherwise $i$ would have received a category-$o$ position instead of a category-$c$ position by non-wastefulness of $C_\rhd^o(\rho;.)=C_{sd}^o(\rho;.)$ by Lemma \ref{basic-lem0}. Moreover,  $C_{sd}^o(\rho;.)$ satisfies no justified envy by Lemma \ref{basic-lem0}. Since $C_\rhd^o(\rho;I)=C_{sd}^o(\rho;I)$, for each $j\in C_\rhd^{o} (\rho;I)$, we have $
	\sigma_j > \sigma_i$ completing the proof that the sequential choice $C_\rhd(\rho;.)$ satisfies compliance with VR protections.
\end{proof*}

\begin{proof*}[Proof of Lemma \ref{basic-lem3}] 
	Fix $\rho\in  \mem$ and $\rho'\in \mem^1$.  Let $C(\rho;.)$ be a choice rule that satisfies non-wastefulness, no justified envy, and compliance with VR protections.    
	$C_{OA}(\rho';.)$ also satisfies these properties by Lemmas \ref{basic-lem1} and \ref{basic-lem2}. Fix a set of individual $I \subseteq \cali$. By definition, $C^o_{OA}(\rho';.)=C^o_{sd}(\rho',.)$ and $C^o_{sd}(\rho',I)  $ consists of $\min\{I, q^o\}$ individuals in $I$ with the highest merit scores. \smallskip
	  
	First, suppose there exists some $i \in  C^o(\rho;I) \setminus C^o_{OA}(\rho';I) $. Thus, $q^o>0$. Then, $i$ is not one of the $\min\{I, q^o\}$ highest-merit-score individuals in $I$. Observe that  if $|I|\le q^o$, then $C^o_{OA}(\rho';I)=I$ contradicting such an individual $i$ exists.  Thus, $|I|>q^o$. Then $\big |C^o_{OA}(\rho';I)\big |=q^o$ and there exists some $j \in C^o_{OA}(\rho';I) \setminus C^o(\rho;I)$ such that $\sigma_j > \sigma_i$. If $j \not \in \h{C}(\rho;I)$ then this together with $\sigma_i <\sigma_j$ and $i \in C^o(\rho;I)$ violate no justified envy property of $C(\rho;.)$ for the open category. Then $j \in C^c(\rho;I)$ for some VR-protected category $c \in \calr$. But then this, together with $\sigma_i <\sigma_j$ and $i \in C^o(\rho;I)$, contradict the fact that $C(\rho;.)$ satisfies compliance with VR protections.   
	This proves such an individual $i$ cannot exist, and therefore, $C^o(\rho;I) \subseteq C^o_{OA}(\rho';I)$. \smallskip
	
	Conversely, suppose there exists some $i \in  C^o_{OA}(\rho';I) \setminus C^o(\rho;I) $. Thus, $q^o>0$. By the first part of the proof, $\big |C^o(\rho;I) \big | < q^o$. If $i \not \in \h{C}(\rho;I)$ then this contradicts non-wastefulness of $C(\rho;.)$ for the open category. Then $i \in C^c(\rho;I)$ for some VR-protected category $c \in \calr$, but this together with $\big |C^o(\rho;I)\big| < q^o$ contradicts the fact that $C(\rho;.)$ satisfies compliance with VR protections.   This proves such an individual $i$ cannot exist, and therefore, with the first part of the proof, we have $C^o(\rho;I) = C^o_{OA}(\rho';I)$.
\end{proof*} 

\subsection{Proofs of Results in Section \ref{sec:EWS-last}} $~$\\ 

\begin{proof*}[Proof of Proposition \ref{prop:NJE-EqualityCode}] Fix $\rho \in \memonezerothree$ that represents the current legislation. Consider its EWS scope-increased profile $\n{\rho} \in \memtwo$. Suppose choice rule $C(\n{\rho};.)$ satisfies no justified envy. Let $I \subseteq \cali$, $i \in I \cap \calj \cap \calf(\rho)$, and $j 
\in I \cap \calj \cap \calm(\rho)$ satisfying 
\[ i \in \h{C}(\n{\rho};I) \; \mbox{ and }  \; j \not\in \h{C}(\n{\rho};I). 
\] 
By the definition of EWS scope increase,
$$
	\{e\}=\rho_i = \n{\rho}_i \subsetneq \n{\rho_j}.
$$
Thus, $i \in C^o(\n{\rho};I) \medcup C^e(\n{\rho};I)$.  Since $j \in \cale^e(\n{\rho}) \medcup \cale^o$, by
no justified envy property of $  C(\n{\rho};.)$ we have $\sigma_i > \sigma_j$. This proves that $C(\n{\rho};.)$ complies with the Equality Code for $I$. Since $I$ is arbitrary, then $C(\n{\rho};.)$ complies with the Equality Code.
\end{proof*} 

\begin{proof*}[Proof of Theorem \ref{thm-EWS}] 
Let $\rho \in \memonezerothree$  represent the current legislation,
and $\n{\rho} \in \memtwo$ be its EWS scope-increased profile.
Let $C\big(\n{\rho};.\big)$ be any choice rule that satisfies non-wastefulness, no justified envy, and compliance with VR protections. Fix $I \subseteq \cali$. If  $\h{C}_{OA}(\rho;I) \setminus \h{C}_{\uline{\rhd}}(\n{\rho};I) =\emptyset$ then the theorem trivially holds, except for the second underbrace statement's proof that is given in the last paragraph of the proof. So suppose
there exists some 
$$
	i \in \h{C}_{OA}(\rho;I) \setminus \h{C}_{\u{\rhd}}(\n{\rho};I).
$$ Since the precedence order of VR-protected categories does not matter in the procedure of $C_{OA}(\rho;.)$, suppose we use $\uline{\rhd}$ in its processing. The procedures of both $C_{OA}(\rho;.)$ and $C_{\uline{\rhd}}(\n{\rho};.)$ coincide until category $e$ is processed. Thus, $i \in C^e_{OA}(\rho;I)$ and 
\begin{align}
	\bigcup_{\nu  \in \calv \setminus \{e\}} C^\nu _{\u{\rhd}}(\n{\rho};I) 
	&=  \bigcup_{\nu  \in \calv \setminus \{e\}} C^\nu _{OA}(\rho;I). \nonumber 
\end{align}
Thus, $i \in C_{OA}^e(\rho;I)  \setminus C_{\u{\rhd}}^e (\n{\rho};I)$. Moreover,  
as $\rho$ represents the current legislation,  $e \in \rho_j \implies j \in \calj \medcap \calf(\rho)$ for every $j \in I$ (including $j=i$). Thus, the first underbrace argument in the statement of the theorem holds, i.e., 
$$
\h{C}_{OA}(\rho;I) \setminus \h{C}_{\u{\rhd}}(\n{\rho};I) = 
	C_{OA}^e(\rho;I)  \setminus C_{\u{\rhd}}^e (\n{\rho};I) \subseteq  \calj  \medcap \calf(\rho).
$$
By Lemma \ref{basic-lem3}, 
\begin{align} C^o(\n{\rho};I) = C_{OA}^o(\rho;I) = C_{\u{\rhd}}^o(\n{\rho};I), \label{eq:W1} 
\end{align} as $C(\n{\rho};.)$ and $C_{\u{\rhd}}^o(\n{\rho};.)$  satisfy non-wastefulness, no justified envy, and compliance with VR  protections (through Lemmas \ref{basic-lem1} and  \ref{basic-lem2} for $C_{\u{\rhd}}(\n{\rho};.)$). Let the set of individuals not chosen by the open category under these three choice rules be
$J= I \setminus C_{OA}^o(\rho;I) $. \smallskip
 
We establish some preparatory relationships to show that $i \notin  C^e(\n{\rho};I)$. To this end, let $c \in \calr^0$.
By no justified envy and non-wastefulness properties, 
$C^c(\n{\rho};I)=C^c_{sd}(\n{\rho};K^c)$ for some $K^c \subseteq J$ by Lemma \ref{basic-lem0-2}. On the other hand, we have   
$C^c_{\u{\rhd}}(\n{\rho};I)=C^c_{sd}(\n{\rho};J)$ by the procedural definition of $C^c_{\u{\rhd}}(\n{\rho};.)$.
By the definition of serial dictatorship and the fact that $J \supseteq K^c$,  
\begin{align}
	\big |C_{\u{\rhd}}^c(\n{\rho};I)\big |&=\big |C_{sd}^c(\n{\rho};J)\big |=\min\Big\{\big |J\medcap \cale^c(\n{\rho})\big |,\;q^c\Big \} \nonumber \\
	&\ge \min\Big\{\big |K^c\medcap \cale^c(\n{\rho})\big |,\;q^c\Big \}= \big |C_{sd}^c(\n{\rho};K^c) \big |=\big |C^c(\n{\rho};I)\big |, \label{eq:W1a}
\end{align}
and additionally, the minimum merit score of individuals in $C_{\u{\rhd}}^c(\n{\rho};I)$ is at least as high as the minimum merit score of individuals in $C^c(\n{\rho};I)$, if they are non-empty. Then,  non-wastefulness and no justified envy properties of $C(\n{\rho};.)$ imply 
\begin{align}
C_{\u{\rhd}}^c(\n{\rho};I) \subseteq \h{C}(\n{\rho};I). \label{eq:W1b}
\end{align}
Since $i \notin  \h{C}_{\u{\rhd}}(\n{\rho};I)$ while $i \in I \medcap \calj \medcap \calf(\rho)\subseteq  I \medcap \cale(\n{\rho})$, we have by the non-wastefulness property of $C_{\u{\rhd}}(\n{\rho};.)$ (established in Lemma \ref{basic-lem1}) that
\begin{align}
	\big |C_{\u{\rhd}}^e(\n{\rho};I)\big |=q^e. \label{eq:W1d}
\end{align}
Moreover, no justified property of  $C_{\u{\rhd}}(\n{\rho};.)$ (also established in Lemma \ref{basic-lem1}) further implies that
\begin{align}
\forall \; j \in C_{\u{\rhd}}^e(\n{\rho};I) \qquad \sigma_j > \sigma_i. \label{eq:W1e}	
\end{align}

We are ready to show that $i \notin C^e(\n{\rho};I)$.  Contrary to the claim, suppose $i \in C^e(\n{\rho};I)$. 
Then no justified envy property of $C(\n{\rho};.)$ and \eqref{eq:W1e} imply 
\begin{align}
	C_{\u{\rhd}}^e(\n{\rho};I) \subseteq \h{C}(\n{\rho};I). \label{eq:W1f}	
\end{align}
However, 
\begin{align*}
	 \sum_{c' \in \calr} \big |C_{\u{\rhd}}^{c'}(\n{\rho};I)  \big | &  = \sum_{c' \in \calr^0} \big | C_{\u{\rhd}}^{c'}(\n{\rho};I)\big | + q^e & \text{by \eqref{eq:W1d}} \\
	 & \ge \sum_{c' \in \calr} \big |C^{c'}(\n{\rho};I)\big | &\text{by \eqref{eq:W1a}\; and \;$\big |C^e(\n{\rho};I)\big |\le q^e$}\\
	& \ge   \sum_{c' \in \calr} \big | C_{\u{\rhd}}^{c'}(\n{\rho};I)\big | + 1 & \text{by \eqref{eq:W1}, \eqref{eq:W1b}, \eqref{eq:W1f}, \; and \; $i \in C^e(\n{\rho};I) \setminus \h{C}_{\u{\rhd}}(\n{\rho};I)$} 
\end{align*}
leading to a contradiction. Thus, $i \notin  C^e(\n{\rho};I)$. \smallskip

Then \eqref{eq:W1} and $i \in \calf(\rho)$ imply $i \notin  \h{C}(\n{\rho};I)$.  We showed that
$$ i \in \h{C}_{OA}(\rho;I) \setminus \h{C}(\n{\rho};I).$$
This concludes the proof of 
$\h{C}_{OA}(\rho;I) \setminus \h{C}_{\u{\rhd}}(\n{\rho};I) \subseteq \h{C}_{OA}(\rho;I)\setminus \h{C}(\n{\rho};I)$. $~$ 
\\

What remains to complete the proof of the theorem is to show the second underbrace expression in its statement.   
By definition, for any $c \in \calr^0, C_{OA}^c(\rho;I)=C_{sd}^c(\rho;J)$, and thus,  $C_{OA}^c(\rho;I)$  consists of the $ \min \Big \{\big |J \medcap \cale^c(\rho)\big |, \; q^c \Big \}$ highest-merit-score individuals in $J \medcap \cale^c(\rho)$. Since $C(\n{\rho};.)$ satisfies no justified envy, then $C_{OA}^c(\rho;I) \subseteq \h{C}(\n{\rho};.)$. Since both choice rules match the same set of individuals to the open category as established before, then the individuals who are chosen under $C_{OA}(\rho;.)$ but not under $C(\n{\rho};.)$ from the set $I$ can be only the ones admitted to category $e$.  Together with the fact that $\rho$ satisfies the current legislation, this implies $$\h{C}_{OA}(\rho;I)\setminus \h{C}(\n{\rho};I) = C_{OA}^e(\rho;I)\setminus \h{C}(\n{\rho};I) \subseteq  \calj \medcap \calf(\rho).$$  
\end{proof*}

Theorem \ref{thm-EWS2} is implied by the following slightly stronger result. 
\begin{theorem} \label{thm-EWS2-general}
Let $\rho \in \memonezerothree$  be a profile of non-overlapping category memberships that represents the current legislation,
and $\n{\rho} \in \memtwo$ be its EWS scope-increased profile. 
Fix a set of applicants $I\subseteq\cali$. 
\begin{enumerate}
	\item If the choice rule $C_{OA}(\rho;.)$ complies with the Equality Code for $I$, then
$  \h{C}_{OA}\big(\rho;I\big) \subseteq \h{C}_{\uline{\rhd}}\big(\n{\rho};I\big)
$. If additionally $\big |I \cap \calj \cap \calf(\rho)\big | \ge q^o+q^e$ holds, then $  \h{C}_{OA}\big(\rho;I\big)= \h{C}_{\uline{\rhd}}\big(\n{\rho};I\big)$.
\item If $  \h{C}_{OA}\big(\rho;I\big)= \h{C}_{\uline{\rhd}}\big(\n{\rho};I\big)$
then $C_{OA}\big (\rho;.\big )$ complies with the Equality Code for $I$. 
\end{enumerate} 
\end{theorem}

\begin{proof*}[Proof of Theorem \ref{thm-EWS2-general}]  Let $\rho \in \memonezerothree$  represent the current legislation and $\n{\rho} \in \memtwo$ be its EWS scope-increased profile. Fix a set of applicants $I\subseteq\cali$. \medskip

\noindent{}\uline{Part (1):} Assume that $C_{OA}(\rho;.)$ complies with the Equality Code for $I$. 
Suppose to the contrary to the claim, there exists some $i \in \h{C}_{OA}(\rho;I) \setminus \h{C}_{\uline{\rhd}}(\n{\rho};I)$. 
Since the precedence order of VR-protected categories does not matter in the procedure of $C_{OA}(\rho;.)$, suppose we use $\uline{\rhd}$ in their processing. Then the procedures of both $C_{OA}(\rho;.)$ and $C_{\uline{\rhd}}(\n{\rho};.)$ coincide until category $e$ is processed. Thus, $i \in C^e_{OA}(\rho;I)$ and 
\begin{align}
	\bigcup_{\nu  \in \calv \setminus \{e\}} C^\nu _{\uline{\rhd}}(\n{\rho};I) 
	&=  \bigcup_{\nu  \in \calv \setminus \{e\}} C^\nu _{OA}(\rho;I).  \label{eq:X1}
\end{align}
Since $\rho$ represents the current legislation,  
\begin{align} 
	i \in I \medcap \calj  \medcap \calf(\rho). \label{eq:X2}
\end{align}
By non-wastefulness of $C_{\uline{\rhd}}(\n{\rho};.)$ (through Lemma \ref{basic-lem1}), 
we have $\big |C^e_{\uline{\rhd}}(\n{\rho};I)\big |=q^e$, as otherwise, $i$ should have been in $C^e_{\uline{\rhd}}(\n{\rho};I)$.
 Then there exists some agent   $j \in C^e_{\uline{\rhd}}(\n{\rho};I) \setminus C^e_{OA}(\rho;I)$. Moreover,
 \begin{align}
 	\sigma_j 
 	&> \sigma_i \label{eq:X3}
 \end{align}
 by no justified envy property of $C_{\uline{\rhd}}(\n{\rho};.)$ for category $e$ (through Lemma \ref{basic-lem1}). As $j \notin  C^e_{OA}(\rho;I)$,  we have by \eqref{eq:X1} $j \notin  \h{C}_{OA}(\rho;I)$. By no justified envy property of $C_{OA}(\rho;.)$ (through Lemma \ref{basic-lem1}),  then   $ e \notin \rho_j$, which satisfies the current legislation. However $e \in \n{\rho}_j$, which implies by the definition of EWS scope increase,
\begin{align} 
	j \in I \medcap \calj  \medcap  \calm(\rho). \label{eq:X4}
\end{align}
Then \eqref{eq:X2}, \eqref{eq:X3}, and \eqref{eq:X4} contradict that $\h{C}_{OA}(\rho;.)$ complies with the Equality Code for $I$. Hence, such an agent $i$ does not exist, completing the proof of $\h{C}_{OA}(\rho;I) \subseteq  \h{C}_{\uline{\rhd}}(\n{\rho};I)$. \medskip

For the second part of Part (1), assume additionally that
\begin{align}
	\big |I \cap \calj \cap \calf(\rho)\big | &\ge q^o+q^e \label{eq:ass}
\end{align}  
holds.
 The first part of the proof implies   $\h{C}_{OA}(\rho;I) \subseteq  \h{C}_{\uline{\rhd}}(\n{\rho};I)$. Suppose $\h{C}_{OA}(\rho;I) \subsetneq  \h{C}_{\uline{\rhd}}(\n{\rho};I)$. Then by \eqref{eq:X1},
$C^e_{OA}(\rho;I) \subsetneq  C^e_{\uline{\rhd}}(\n{\rho};I)$. Hence $\big |C^e_{OA}(\rho;I)\big |<q^e$. But then \eqref{eq:ass} implies that there exists some $j \in I \medcap \cale^e (\rho)$ such that $j \notin \h{C}_{OA}(\rho;I)$, contradicting non-wastefulness (through Lemma \ref{basic-lem1}).   Thus, $\h{C}_{OA}(\rho;I) =  \h{C}_{\uline{\rhd}}(\n{\rho};I).$ \medskip

\noindent{}\uline{Part (2):} Suppose $\h{C}_{OA}\big(\rho;I\big) = \h{C}_{\uline{\rhd}}\big(\n{\rho};I\big)$. By Lemma \ref{basic-lem1},  $\h{C}_{\uline{\rhd}}\big(\n{\rho};.)$ satisfies no justified envy. Thus, by Proposition \ref{prop:NJE-EqualityCode},  $\h{C}_{\uline{\rhd}}\big(\n{\rho};.)$ satisfies compliance with the Equality Code for $I$. Then, $\h{C}_{OA}\big(\rho;.)$  also satisfies compliance with the Equality Code for $I$. 
\end{proof*}

\subsection{Proofs of Results in Section \ref{sec:EWS-first}} $~$\\

\begin{proof*}[Proof of Theorem \ref{thm-VR-caste}] Let $\rho \in \memonezerothree$ represent the current legislation. Consider its EWS scope-elevated profile $\n{\rho} \in \memtwo$. Fix also $\overline{\rhd} \in \Delta^{o,e}$.  
Let choice rule $C(\n{\rho};.)$  respect mobility from reparatory categories to EWS  and  satisfy non-wastefulness, 
no justified envy, and compliance with VR protections.

Let $I \subseteq \cali$.
By Lemma \ref{basic-lem3},   $C^o(\n{\rho},I)=C^o_{sd}(\rho,I)$ and $C_{\overline{\rhd}}^o(\n{\rho};I)=C^o_{sd}(\rho,I)$. Thus,
	\begin{align}
		C^o(\n{\rho},I)=C_{\overline{\rhd}}^o(\n{\rho};I). \label{eq:thm-VR-1}
	\end{align}
	Let 
		$J = I\setminus C^o(\n{\rho},I)$.  	
We prove the following claim:

 \begin{claim}
 	\label{cl:thm-VR-caste-based} $C^{e}(\n{\rho},I)=C_{\overline{\rhd}}^{e}(\n{\rho};I).$
\end{claim}
 
\begin{proof}[Proof of Claim \ref{cl:thm-VR-caste-based}] 
	Recall that $C_{\overline{\rhd}}^{e}(\n{\rho};I)=C_{sd}^{e}(\n{\rho};J)$. Let $K^e=J \setminus \bigcup_{c \in \calr^0} C^c(\rho;I)$.  Since $C(\n{\rho};.)$ satisfies non-wastefulness and no justified envy, by Lemma \ref{basic-lem0-2} we have $C^{e}(\n{\rho};I)=C_{sd}^{e}(\n{\rho}; K^e)$. 
	 To show that $C^{e}(\n{\rho};I)=C_{sd}^{e}(\n{\rho};J)$, we assume to the contrary to the claim there exists some $i \in C_{sd}^{e}(\n{\rho};J) \setminus C^{e}(\n{\rho};I).$ Then there exists some $j \in  C^c(\n{\rho};I) \medcap C_{sd}^{e}(\n{\rho};J)$ for some $c \in \calr^0$, as otherwise, $C_{sd}^{e}(\n{\rho};J)=C_{sd}^{e}(\n{\rho};K^e)=C^{e}(\n{\rho};I)$. Thus, $j \in I \medcap \calj \medcap \calm(\rho)$. 
	 	Moreover, $j$ is one of the $\min \Big \{\big |J \medcap \cale^e(\n{\rho}) \big |, \; q^e \Big \}$ highest merit-score individuals in $J \medcap \cale^e(\n{\rho})$ as $j \in C_{sd}^{e}(\n{\rho};J)$. Thus, as $j 
	 	\notin C^e(\n{\rho};I)$, either $\big |C^e(\n{\rho};I)\big |<q^e$ or there exists some $i \in  C^e(\n{\rho};I)$ such that $\sigma_i > \sigma_j$. Both of these contradict that $C(\n{\rho};.)$ maintains the elevated status of caste-based VR protections. Thus, such an individual $j$ cannot exist, and therefore, $C^{e}(\n{\rho};I)=C_{sd}^{e}(\n{\rho};J)$. Thus,
	 	$$C^{e}(\n{\rho};I)=C^{e}_{\overline{\rhd}}(\n{\rho};I).$$
	\end{proof}
  Let 
  $$
  	K=J \setminus C^{e}(\n{\rho};I).
  $$
  Let $c \in \calr^0$. 
  Since both categories $o$ and $e$ precede all categories in $\calr^0$, precedence order of categories in $\calr^0$ is immaterial to the outcome of $C_{\overline{\rhd}}(\n{\rho};.)$. In particular,  $C_{\overline{\rhd}}^{c}(\n{\rho};I)=C_{sd}^c(\n{\rho};K)$. 
  
We show $C^c(\n{\rho};I)=C_{sd}^c(\n{\rho};K)$.   
  By Lemma \ref{basic-lem0-2}, as $C(\n{\rho};.)$ satisfies non-wastefulness and no justified envy, $C^c(\n{\rho};I)=C_{sd}^c(\n{\rho};K^c)$ where $K^c=K \setminus \bigcup_{{c'} \in \calr^0 \setminus \{c\}} C^{c'}(\n{\rho};I)$.  Since $\rho$ represents the current legislation, under $\n{\rho}$, no VR-protected categories in $\calr^0$ overlap. Thus, no individual in $K\cap \cale^c (\n{\rho})$ is selected by a category in $\calr^0\setminus\{c\}$ and $K \medcap \cale^c(\n{\rho}) \subseteq K^c$. Hence, $C_{sd}^c(\n{\rho};K)=C_{sd}^c(\n{\rho};K^c)$ implying $$C^c(\n{\rho};I)=C_{\overline{\rhd}}^c(\n{\rho};I).$$
  This together with Eq. \eqref{eq:thm-VR-1} and Claim \ref{cl:thm-VR-caste-based} imply that $C(\n{\rho};I)=C_{\overline{\rhd}}(\n{\rho};I)$.
\end{proof*}

\subsection{Proofs of Results in Section \ref{sec:mOA}} $~$\\
	
\begin{proof*}[Proof of Proposition \ref{prop-mOA}] 
Fix a profile of category memberships $\rho \in \mem$ and a set of individuals  $I \subseteq \cali$. Let $\cale^c=\cale^c(\rho)$ for each $c \in \calr$. \smallskip

	\textit{Non-wastefulness:} Suppose $v \in \calv$ be such that  $\big |C_{\o}^{v} (\rho;I) \big |<q^{v}$,
	and there exists some $j\in I\setminus\aco (\rho;I)$. If $v=o$, then this contradicts $j$ not being selected in Step 1 of the meritorious O\&A choice rule, which is the category-$o$ serial dictatorship. Suppose $v \in \calr$. Thus, even though Step 2 of the procedure leaves vacant positions in category $v$, $j$ does not receive a position, meaning that she cannot increase the VR-utilization of $\bigcup_{c \in \calr} C_{\o}^c(\rho;I)$. Then $j \not \in \cale^v$. We showed $C_{\o} (\rho;.)$ is non-wasteful. \smallskip

	\textit{No justified envy:} Let $v \in \calv$ and suppose there exists $j \in I \setminus \h{C}_{\o}(\rho;I)$. We have two cases:
	
	(1) Consider the case $v=o$. By construction of Step 1 of the meritorious O\&A choice rule, if  $i \in C_{\o}^o(\rho;I)=C^o_{sd}(\rho;I)$, then $i$ is chosen instead of $j$,  in the category-$o$ serial dictatorship. Hence, $\sigma_i > \sigma_j$.  
	
	(2) Consider the case $v \in \calr$. Suppose $i \in C_{\o}^v(\rho;I)$. Then $i \in \cale^v$. Let  
 	 $J=I \setminus C_{\o}^o(\rho;I)$. Observe that Step 2 of the meritorious O\&A choice rule is the greedy algorithm for matroid $\big (J,\cala(q',J) \big) $ where the auxiliary reservation vector $q' \in \calq$ is obtained from $q$ by setting the open category positions to zero and keeping every other category's positions the same as in $q$ (see Remark \ref{rem-mOA}). Hence, in Step 2, the greedily assignable subset $G(q',J;\sigma)$ is the set of individuals chosen. In particular, $i \in G(q',J;\sigma)$.
 	 
 	 Suppose, for a contradiction, $\sigma_j > \sigma_i$ and $j \in \cale^v$. 
 	  	 
 	 We first show that set $K$ defined as $K=\big( G(q',J;\sigma) \setminus \{i\} \big) \cup \{j\}$ is an assignable subset of $J$ under $q'$:  To see this, pick a greedy matching $\mu \in \bbm(q',J)$ assigning $i$ to $v$.  Such a greedy matching exists as  $i \in C_{\o}^v(\rho;I)$ (see Remark \ref{rem-mOA}). By definition of a greedy matching, $J_{\mu}=G(q',J;\sigma)$. Since both $i,j\in \cale^v$, we can modify $\mu$ by assigning $j$ to $v$ instead of $i$, leaving $i$ unmatched, and keeping all other assignments the same. Thus, we obtain a matching in $\bbm(q',J)$ that matches all individuals in $K$, showing that $K\in \cala(q',J)$. 
 	 
 	  Since $\sigma_j > \sigma_i$,  $j$ is processed before $i$ in the process of the greedy algorithm. 
 Then $j$ should be chosen instead of $i$, as $K$ is an assignable subset of $J$ under $q'$, including $j$ and all previously committed individuals in the greedy algorithm before $j$. This contradicts $j \not \in  G(q',J;\sigma) \subseteq \h{C}_{\o}(\rho;I)$. Hence, $\sigma_i > \sigma_j$ or $j \not \in \cale^v$. 
 	 
 	 Thus, $C_{\o}(\rho;.)$ satisfies no justified envy.   \smallskip

\textit{Compliance with VR protections:} Let $c\in \calr$ and $i\in C_{\o}^{c} (\rho;I)$.
	In executing the procedure of the meritorious O\&A choice rule,  the open category is processed first, and individual $i$ is still available when the VR-protected categories are about to be processed in Step 2. Thus, it should be the case that   $\big |C_{\o}^{o} (\rho;I)\big |=q^{o}$, as otherwise $i$ would have received a category-$o$ position instead of a category-$c$ position. Moreover,  $C_{\o}^o(\rho;I)$, which is formed by a serial dictatorship, includes the highest $\sigma$-score $q^o$ individuals from $I$, implying that for every $j \in C_{\o}^{o} (\rho;I)$, $\sigma_j > \sigma_i$.
	Thus,  $C_{\o}(\rho;.)$ satisfies compliance with VR protections. \smallskip
	
	Additionally, for the final part of the proposition, assume that $\rho \in \memonezerothree$ represents the current legislation. Consider its EWS scope-elevated profile $\n{\rho} \in \memtwo$. Since $\co(\n{\rho},.)$ satisfies no justified envy, by Proposition \ref{prop:NJE-EqualityCode}, it also complies with the Equality Code. 
	\end{proof*}

\begin{proof*}[Proof of Theorem \ref{thm-mOA}] 
	Fix $I \subseteq \cali$. By Lemma \ref{basic-lem3}, for any non-overlapping profile of category memberships $\rho' \in \mem$ we have $C_{\o}^o(\rho;I)=C^o_{sd}(\rho';I)$ and $C^o(\rho;I)=C^o_{sd}(\rho';I)$, implying that	
	\begin{align} 
		C_{\o}^o(\rho;I) = C^o(\rho;I). \label{eq:EWS-smart-1}
	\end{align}
	Let $J=I\setminus  C_{\o}^o(\rho;I)$. 
	
	Define the auxiliary reservation vector $q'\in \calq$ such that the open category has no position and each VR-protected category has the same number of positions as it has under $q$: ${q'}^o=0$ and ${q'}^c=q^c$ for each $c\in \calr$.  	Construct the matching $\mu \in \bbm(q',J)$ by assigning  positions  of each category $c\in \calr $ under $\mu$  to  the individuals in $C^c(\rho;I)$. Thus, the set $\bigcup_{c \in \calr}C^c(\rho;I)$ is an assignable subset of $J$ under $q'$, i.e.,  
	\begin{align}
		\bigcup_{c \in \calr}C^c(\rho;I) \in \cala(q',J). \label{eq:EW-smart-2}
	\end{align}
	Step 2 of the meritorious O\&A rule is the greedy algorithm for matroid $\big (J,\cala(q',J) \big) $ under $\sigma$ (see Remark \ref{rem-mOA}), 
 implying that 	
	\begin{align*}
		\bigcup_{c\in \calr}C_{\o}^c(\rho;I)=G(q',J;\sigma). 	
	\end{align*}
	Thus, by Lemma \ref{lem-Gale} and Eq. \eqref{eq:EW-smart-2}, $\bigcup_{c\in \calr}C_{\o}^c(\rho;I)$ Gale dominates $\bigcup_{c \in \calr}C^c(\rho;I).$
	This statement and Eq. \eqref{eq:EWS-smart-1} show that set $\h{C}_{\o}(q;I)$ Gale dominates set $\h{C}(q;I)$.
\end{proof*}
 
\end{appendix}
\end{document}